\documentclass[journal]{IEEEtran}

\usepackage{array}
\usepackage[utf8]{inputenc}
\usepackage{amsmath, commath}
\usepackage{amsthm}
\usepackage{amsfonts}
\usepackage{amssymb}
\usepackage{bm}
\usepackage{bbm}
\usepackage{graphicx}
\usepackage{graphics}
\usepackage{xcolor}
\usepackage{float}
\usepackage{tikz}
\usetikzlibrary{shapes, snakes, patterns}

\usepackage{epstopdf}
\usepackage[export]{adjustbox}
\usepackage[font={footnotesize}]{caption}
\usepackage[list=true]{subcaption}
\usepackage{color}
\usepackage{algorithm}
\usepackage{algorithmic}

\newtheorem*{example}{Example}
\usepackage{enumerate} 
\usepackage[mode=buildmissing]{standalone}

\usepackage{balance}
\usepackage{cite}
\usepackage{suffix}
\usepackage{mathtools}
\usepackage{tabularx, booktabs}
\usepackage{multirow}
\usepackage{diagbox}
\usepackage{amsfonts}
\usepackage{amssymb}
\usepackage{textcomp}
\usepackage{dsfont} 
\usepackage{subcaption}
\usepackage{pgfplots,pgfplotstable}
\usepackage{bm}
\usepackage{physics}
\usepackage{svg}
\usepackage{gensymb}
\usepackage{url}
\usepackage{tikz}
\begin{document}
\title{End-to-End Learning for VCSEL-based Optical Interconnects: State-of-the-Art, Challenges, and Opportunities}
\author{
    Muralikrishnan Srinivasan, Jinxiang Song, \emph{Student Member, IEEE,} Alexander Grabowski, \emph{Member, IEEE,}
    Krzysztof Szczerba, \emph{Senior Member, IEEE,}
    Holger K. Iversen, Mikkel N. Schmidt,
    Darko Zibar,
 Jochen Schr\"oder, \emph{Member, IEEE, Senior Member, OSA,}
 Anders Larsson, \emph{Fellow, IEEE, Fellow, OSA}, Christian H\"ager, \emph{Member, IEEE,} and Henk Wymeersch, \emph{Senior Member, IEEE}
\thanks{
This work was supported by the Swedish Foundation for Strategic Research (SSF, HOT-OPTICS Project).  The work of J.~Song was supported by the Knut and Alice Wallenberg Foundation, grant No.~2018.0090. The work of C.~Häger was also supported by the Swedish Research Council under grant no. 2020-04718. D.~Zibar was supported by the European Research
Council through the ERC-CoG FRECOM project under Grant Agreement
771878 and Villum Synergy grant \emph{(Corresponding author: Muralikrishnan Srinivasan.)}}
\thanks{Muralikrishnan Srinivasan, Jinxiang Song, Christian~H\"ager, and Henk Wymeersch are with the Department of Electrical Engineering, Chalmers University of Technology, Gothenburg 41296, Sweden
(e-mails: \{mursri, jinxiang,   christian.haeger, henkw\}@chalmers.se).
}
\thanks{Alexander Grabowski, Jochen Schröder, and Anders Larsson are with the Photonics Laboratory, Department of Microtechnology and Nanoscience (MC2), Chalmers University of Technology, SE-41296 Gothenburg, Sweden
 (emails: \{alexander.grabowski, jochen.schroeder, anders.larsson\}@chalmers.se).
}
\thanks{Krzysztof Szczerba is with OpenLight Photonics, Mountain View, CA 94087, USA (email: kszczerba@openlightphtonics.com).}
\thanks{Darko Zibar and Holger K. Iversen are with the Department of Photonic Engineering, Technical University of Denmark, 2800 Kgs. Lyngby, Denmark (email: dazi@fotonik.dtu.dk)}
\thanks{Mikkel N. Schmidt is with the Department of Applied Mathematics and Computer Science, Technical University of Denmark. 2800 Kgs. Lyngby, Denmark (email: mnsc@dtu.dk).}
}
\maketitle

\begin{abstract}
    Optical interconnects (OIs) based on vertical-cavity surface-emitting lasers (VCSELs) are the main workhorse within data centers, supercomputers, and even vehicles, providing low-cost, high-rate connectivity. VCSELs must operate under extremely harsh and time-varying conditions, thus requiring adaptive and flexible designs of the communication chain. Such designs can be built based on mathematical models (model-based design) or learned from data (machine learning (ML) based design). Various ML techniques have recently come to the forefront, replacing individual components in the transmitters and receivers with deep neural networks. Beyond such component-wise learning, end-to-end (E2E) autoencoder approaches can reach the ultimate performance through co-optimizing entire parameterized transmitters and receivers. 
    This tutorial paper aims to provide an overview of ML for VCSEL-based OIs, with a focus on E2E approaches, dealing specifically with the unique challenges facing VCSELs, such as the wide temperature variations and complex models. 
\end{abstract}


\section{Introduction}

Since the 1980’s fiber optics has been deployed at a massive scale in communication networks, from high-capacity transoceanic links and metropolitan networks to broadband fiber-to-the-home connections~\cite{agrawal2016}. Fiber optical communication forms the backbone of the network that carries the global Internet traffic. Because of the long reach of these optical links (from several to thousands of km's), single-mode fiber (SMF) is exclusively used as the transmission medium. Wavelengths are typically in the O-band (around 1310 nm, the zero-dispersion wavelength) for short distances or in the C-band (around 1550 nm, the minimum attenuation wavelength) for long-haul communications.
With the emergence of the Internet, data centers were established to store, process, and distribute data between service providers and users. Cloud services currently provide a myriad of information technology  resources distributed over the global network. According to the Cisco Visual Networking Index (VNI) for 2017-2022, in the year 2017, 1.5 ZB of Internet Protocol (IP) traffic was transmitted over the Internet, with projections of 4.8 ZB traffic for the year 2022 
\cite{ciscovni}. However, the amount of data center traffic is even higher and amounted to 20 ZB in 2021, with more than 70\% (15 ZB) staying within the data center~\cite{ciscoforecast}. To sustain this traffic, a high-capacity network is needed to interconnect compute, storage, and switch units. Therefore, optical interconnects (OIs) have, to a large extent, replaced copper-based interconnects in datacenter and high-performance computing environments. 

\begin{figure}
    \centering
    \includegraphics[width=\columnwidth]{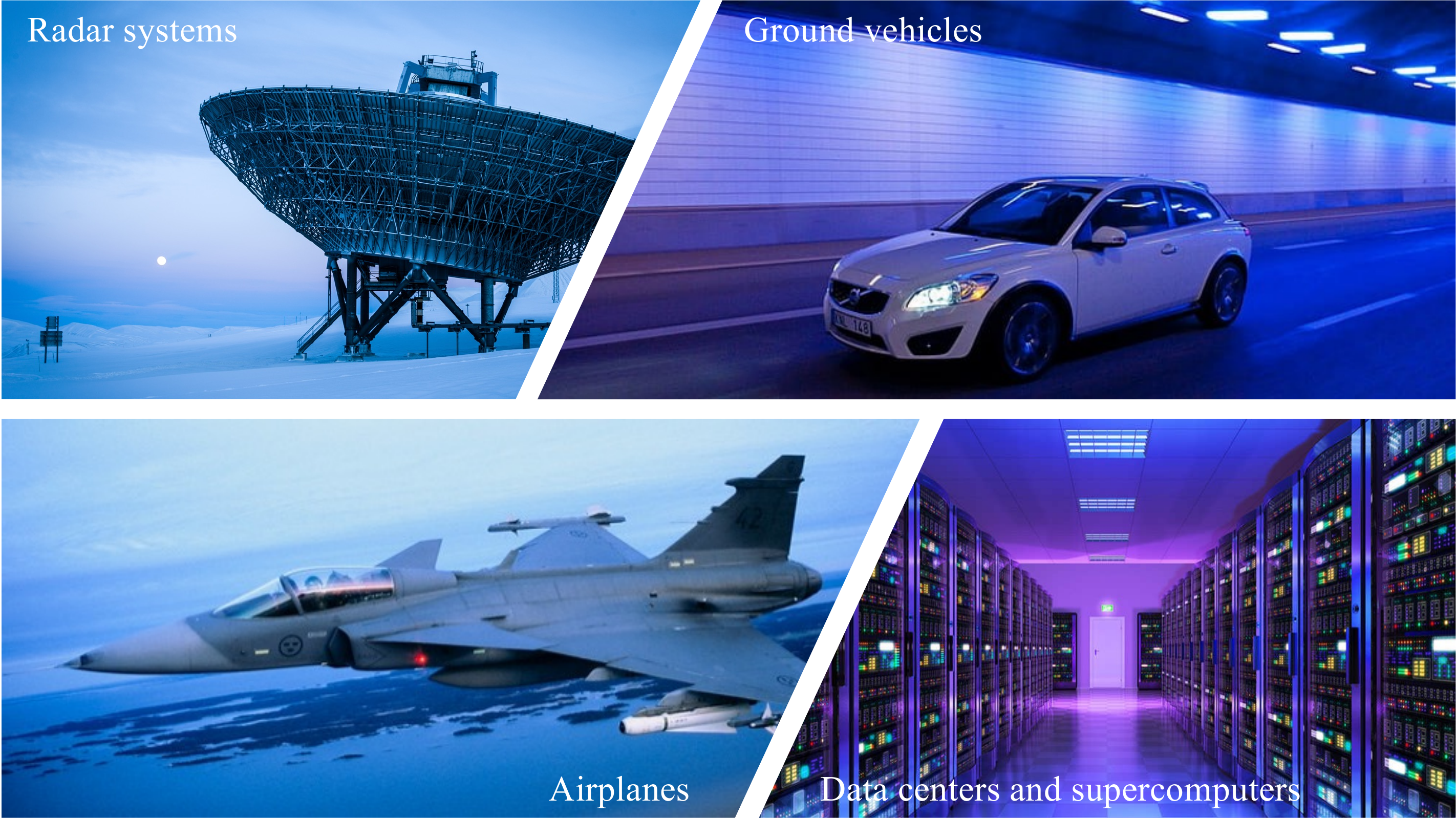}
    \caption{A wide variety of applications rely on VCSELs and operate in extremely harsh environments.} \vspace{-3.2mm}
    \label{fig:applications}
\vspace{-11pt} \end{figure}

\subsection{The Rise of VCSEL-based OIs}
Within the data center, different optical technologies are used at different levels of the network to provide the capacity needed at a minimum of cost and power consumption~\cite{cheng2018}. At the higher levels, with distances of 100-300 m and above, depending on the symbol rate, SMF is used together with O-band externally modulated lasers or silicon photonics transceivers. At the lower level, with a reach of 30-300 m and below, again depending on the symbol rate, multimode fiber (MMF) is used together with 850 nm vertical-cavity surface-emitting lasers (VCSELs). At this level, cost and power efficiency are of utmost importance because of the large number of interconnects. The VCSEL is the most cost and power-efficient light source~\cite{maharry202150}. It can be directly modulated at high speed and enables the smallest footprint transceivers~\cite{mahgerefteh2016}.
Over the last decade, the modulation bandwidth of 850 nm VCSELs has increased from 20 to 30+ GHz~\cite{cheng2022}. This has enabled VCSEL-based short-reach OIs with lane rates of 56 Gbps (e.g., 400GbE SR8, InfiniBand HDR), using Pulse-Amplitude Modulation 4-level (PAM4) modulation at 28 Gbaud~\cite{nvidia}. Lane rates of 112 Gbps (e.g., 800GbE SR8, InfiniBand NDR), again using PAM4 (56 Gbaud), will follow, enabled by commercially available VCSELs with bandwidth approaching 30 GHz~\cite{murty2021,hoser2022}. Such interconnects, with superior power and cost efficiency and operating at 850 nm, use multimode VCSELs, MMF, and large-size photodetectors for tolerance to misalignment, which significantly reduces cost as it allows for passive alignment during transceiver assembly. Further developments of VCSELs may enable lane rates of 224 Gbps (e.g., InfiniBand XDR at 112 Gbaud PAM4) and aggregate transceiver capacities beyond 1 Tbps. VCSEL revenues for datacom are expected to exceed 2 BUSD in 2027~\cite{yole}. In comparison, the optical datacom transceiver market is expected to grow from 5.9 BUSD in 2021 to 16.8 BUSD in 2027, representing a 19\% annual growth rate~\cite{yole2}. 

Optical transceivers are pluggable, which means they are attached to the front panel of, e.g., a server or switch. With increasing compute and switch capacity, and therefore increasing interconnect bandwidth, the electrical interconnects transporting data from the integrated circuits (ICs) on the circuit boards to the optical transceivers on the front panel cannot provide the bandwidth or the low loss transmission needed and the optical transceiver port count and density becomes excessive. Therefore, the transceivers must migrate into the units close to the compute or switch ICs. This is referred to as co-packaged optics~\cite{minkenberg2021} and may require the transceivers to operate at significantly higher temperatures. This represents a much harsher environment, which creates challenges for high data-rate operations. 
%
%
Similar needs for and requirements on VCSEL-based transceivers, in terms of, e.g., capacity, power consumption, and temperature, are found in high-performance computing systems (supercomputers)~\cite{rumley2017}. In fact, there is a data center-supercomputer convergence as (big) data is increasingly processed in data centers using artificial intelligence (AI) and deep learning tools~\cite{hoefler2022}. This requires low interconnect latency, in addition to high interconnect capacity, high interconnect density, and low power consumption.
Finally, short-reach OIs are considered for other types of networks to enable higher interconnect capacity. For example automotive optical networking which may develop into the next large-volume application for datacom VCSELs. Automotive networking represents another harsh environment application, with an operating temperature from -40 to +125$\degree$C. VCSELs for this application are under development \cite{king2021, aoki2021} and VCSEL-based networking standards are developed in the IEEE P802.3cz Multi-Gigabit Optical Automotive Ethernet Task Force for lane rates up to 50 Gbps \cite{ieeereport}. Optical networking in radars and other military/defense systems is yet another harsh environment application for which rugged, high-capacity VCSEL-based transceivers are developed \cite{amphenolreport}.

\subsection{VCSEL-based OIs: from Model-based to ML-based Designs}


\begin{figure*}
    \centering
    \includegraphics[width=\linewidth]{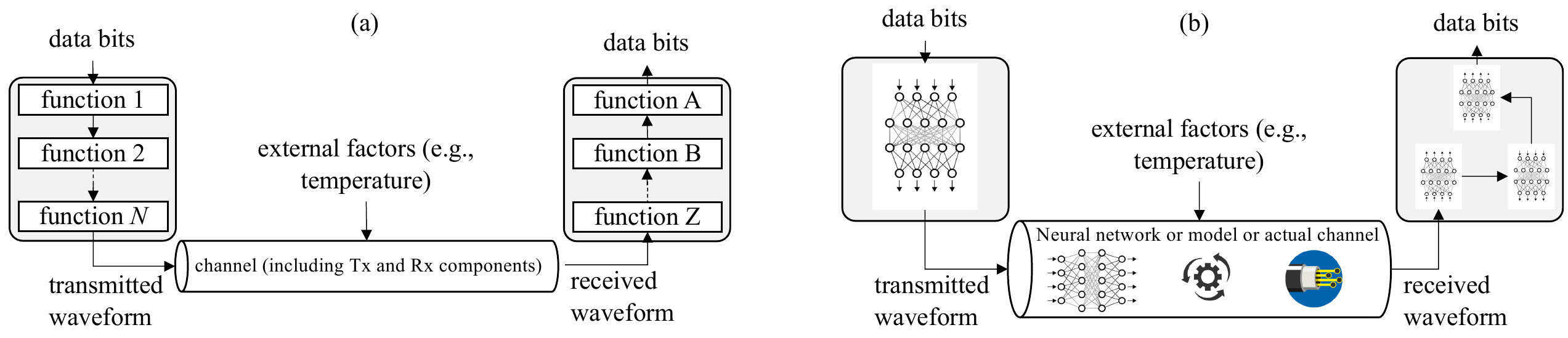}
    \caption{(a) A conventional approach to VCSEL-based fiber-optical communication, in which several functions (e.g., encoding, modulation, pulse shaping) at the transmitter (Tx) convert incoming bits to a waveform, while the receiver (Rx) applies functions (e.g., synchronization, equalization, demodulation) to recover the data. (b) An E2E ML perspective, where transmitter and receiver are replaced by NNs (in case of the receiver also with a functional decomposition), and where the channel may also  be, in part, modeled by a NN. The entire system is learned E2E via a suitable loss function. }\vspace{-5mm}
    \label{fig:AEoverview}
\end{figure*}

These considerations of cost, energy efficiency, and temperature variations have several implications in terms of communication capacity. A model of the end-to-end (E2E) channel must be available for a communication system to operate close to its theoretical capacity. Modeling individual components is already very challenging, let alone modeling the E2E channel. Secondly, even if a model is available, it would be overly complex, involving the concatenation of many nonlinear, frequency-selective, and noisy sub-models, which in turn precludes the possibility of designing an optimal transmitter and the corresponding optimal receiver. Finally, the severe temperature swings demand both transmitter and receiver adaptivity to select the best modulation format, code rate, and waveform parameters to maximize data throughput from the lowest to the highest temperature. 

\emph{Model-based communication} systems are built on the combination of two principles---\emph{functional decomposition} and \emph{interface specification}---in order to tame the complexity and thus cost and energy consumption.  The functional decomposition (see Fig.~\ref{fig:AEoverview}-(a)) involves both the separation into layers (e.g., physical layer and access control layer) as well as the separation within a layer. In the physical layer, functional decomposition involves separating source coding, channel coding, modulation, pilot embedding and waveform selection at the transmitter and their inverse operations filtering, synchronization and channel estimation, equalization, demodulation, and decoding at the receiver. Complementary to the functional decomposition is the interface specification, which determines the input and output of each function, thereby connecting it coherently with other functional blocks. 
The combination of functional decomposition and interface specification has been successfully applied to various wired and wireless communication systems, provided the channel model is not overly complicated. Only for specific channels, including the additive white Gaussian noise (AWGN) channel, a truly optimal functional decomposition in terms of capacity is possible. For most real channels, each functional block is optimized separately, often based on a local criterion, in the hope that good E2E performance can be achieved in terms of rate, latency, etc.

\emph{Machine learning (ML)} provides an attractive alternative to traditional model-based approaches to overcome the three challenges (modeling, design, and adaptivity) \cite{zhong2018digital}. In particular, deep ML, which harnesses the flexibility and massive over-parametrization to perform classification, regression, clustering, or rewards-driven action-taking, has been shown to outperform most man-made approaches in nearly all engineering fields, including language and image processing and games. 
Models of an E2E system can be inferred using the data obtained from the communication system. From classical models of components, neural network (NN) equivalents can be constructed and refined using data and then plugged into learning methods to optimize communication parameters or algorithms. Receiver-side algorithms for equalization, synchronization, data detection, and decoding can be learned either by mimicking conventional algorithms and then further optimizing performance or using a deep neural network (DNN) from scratch \cite{shen2011fiber, gaiarin2016high, ge2020compressed,deligiannidis2020compensation,argyris2018photonic}. Adaptivity can be built into these learned models and methods, for example, by providing suitable inputs from the external environment, which describes the state of the outside world. Such inputs can comprise the temperature but can also be any input from which the ML model can learn suitable embeddings and remove non-salient information. Transmitter-side methods for encoding, modulation, pre-distortion, and waveform synthesis are generally harder to learn as they require a matching receiver \cite{schaedler2019ai}. 


Within the field of (deep) ML, a paradigm shift occurred with the introduction of E2E learning, driven mainly by the use of autoencoders (AE) and alternating optimization \cite{o2017introduction}. An AE (see Fig.~\ref{fig:AEoverview}-(b)) mimics a communication system, in that it comprises an encoder NN (the transmitter, converting bits to waveforms), a bottleneck (the propagation channel), and a decoder NN (the receiver, converting the received waveform back to bits). When differentiable channel models are available (possibly in the form of a NN), the encoder and decoder can be jointly optimized, by propagating gradients of the loss function from the decoder, over the bottleneck, through the encoder. With sufficient training and optimization of hyperparameters, such E2E optimization has been successful in a wide variety of applications~\cite{Doerner2018,stark2019joint, raj2018backpropagating,Ye2018}, including in fiber-optical communications \cite{li2018achievable, Karanov2018, Jones2018, Karanov2019end,Uhlemann2020, Jovanovic2021}. 
Nevertheless, the AE has two drawbacks: (i) it does not rely on the principle of functional decomposition and (ii) it requires a differentiable channel model. The former drawback can be addressed by decomposing the encoder and decoder into smaller NNs and constraining their interface based on the conventional model-based designs. This provides a means to perform E2E learning with a clearly identifiable modulator, waveform generator, equalizer, decoder, etc. The second drawback is addressed by either first learning a model of the channel and/or components, so that a differentiable NN can be used as a proxy of the real non-differentiable channel, or by using so-called gradient free methods, such as reinforcement learning. 

In summary, while E2E learning has the potential to learn an optimal communication transmitter and receiver without any a priori model, it can benefit significantly by providing a functional decomposition and from model-based knowledge. Therefore, it is natural that OIs can benefit from using ML techniques, especially E2E learning techniques, given the rich body of literature providing model-based knowledge on VCSELs. However, given the harsh operating conditions of VCSELs, and the complexity of modeling VCSELs, the E2E techniques applied to conventional intensity-modulation/direct-detection (IM/DD) links cannot be directly applied to VCSEL-based OIs. Therefore, reviewing the characteristics of VCSEL and the challenges in ML-based VCSEL modeling becomes imperative. Then, a comprehensive review of the existing ML-based approaches in designing specific components of OIs is necessary. Finally, it is crucial to explore and study the opportunities in E2E learning in the context of VCSEL-based OIs.

\subsection{Paper Contributions}
There have been several reviews and surveys on ML for optical networking \cite{musumeci2018overview}, for recent developments of short-reach optical communications \cite{chagnon2019optical}, and for short-reach communications \cite{xie2022machine}. Unlike \cite{chagnon2019optical} and \cite{xie2022machine}, we do not focus on short-reach communications but on VCSEL-based OIs, which bring several modeling challenges not found in conventional IM/DD systems. 
Unlike \cite{khan2019optical, pan2021machine, saif2020machine}, our tutorial's primary focus is not on component-specific learning, such as nonlinearity compensation or optical performance monitoring but also extends to E2E learning using AEs. 
Finally, beyond being a survey that summarizes state-of-the-art, this tutorial is also intended for researchers in optical communication who require some demonstration and insights to model O/E components using ML and utilize them in the context of E2E learning.  
Our contributions are as follows:
\begin{enumerate}
    \item \textbf{Introduction of VCSELs and ML-based modeling of their dynamics:} We present a substantial introduction to VCSELs, including circuit level models, followed by a detailed study on challenges in modeling VCSEL dynamics. We then present Volterra series and NN-based methods to construct so-called surrogate models, which can capture VCSEL characteristics with controllable fidelity. These models form an essential building block in E2E AE-based designs.
    \item \textbf{Review of ML-based pre- and post-compensation methods:}
    We comprehensively review ML-based equalization and digital pre-distortion/nonlinearity compensation techniques. We present compensation techniques that can work in the presence and absence of surrogate models of the VCSEL, and show how ML-based methods can generally match and sometimes outperform standard model-based approaches. 
    \item \textbf{In-depth discussion on AEs for VCSEL-based OIs:}
    Most importantly, we present the state-of-the-art AE-based E2E techniques used in fiber-optic communications, especially in conventional IM/DD systems and coherent systems. We comprehensively discuss the methodology, the challenges, and the opportunities, like temperature-adaptive transceivers, and gradient-free approaches, in AE-based E2E learning for VCSEL-based OIs. 
\end{enumerate}

The remainder of this tutorial is organized as follows. In Section \ref{sec:vcselintro}, we provide an introduction to VCSELs and their dynamic. Then, in Section \ref{sec:modeling}, we study different surrogate models, including circuit models, Volterra series models,  and NN-based models. In Section \ref{sec:Component}, we review pre- and post-compensation of individual components, and show how ML-based methods can be applied to replace standard model-based counterparts. The concepts from these earlier sections are then integrated into Section \ref{sec:AutoEncoder}, which deals with AEs. We show how the important challenges in VCSEL-based IOs can be addressed through E2E learning. Finally, Section \ref{sec:Conclusions} wraps up the tutorial with our main conclusions and opportunities for future research in this emerging area. 



\section{Introduction to VCSELs}\label{sec:vcselintro}
In this section, we introduce the history of VCSELs, the basic structure of a VCSEL, and provide an introduction to its fundamental operating characteristics. We end the section with a discussion on the importance of modeling VCSELs. 

\subsection{VCSELs: History and Main Properties}
Demonstrated in 1979 \cite{Soda_1979}, the VCSEL has since become an attractive choice for applications, ranging from OIs \cite{LarssonAdvances}, to sensing and high-power applications (VCSEL arrays) \cite{VCSELarraysLidar}. 
\begin{figure}[h]
\centering
\includegraphics[width=0.9\linewidth]{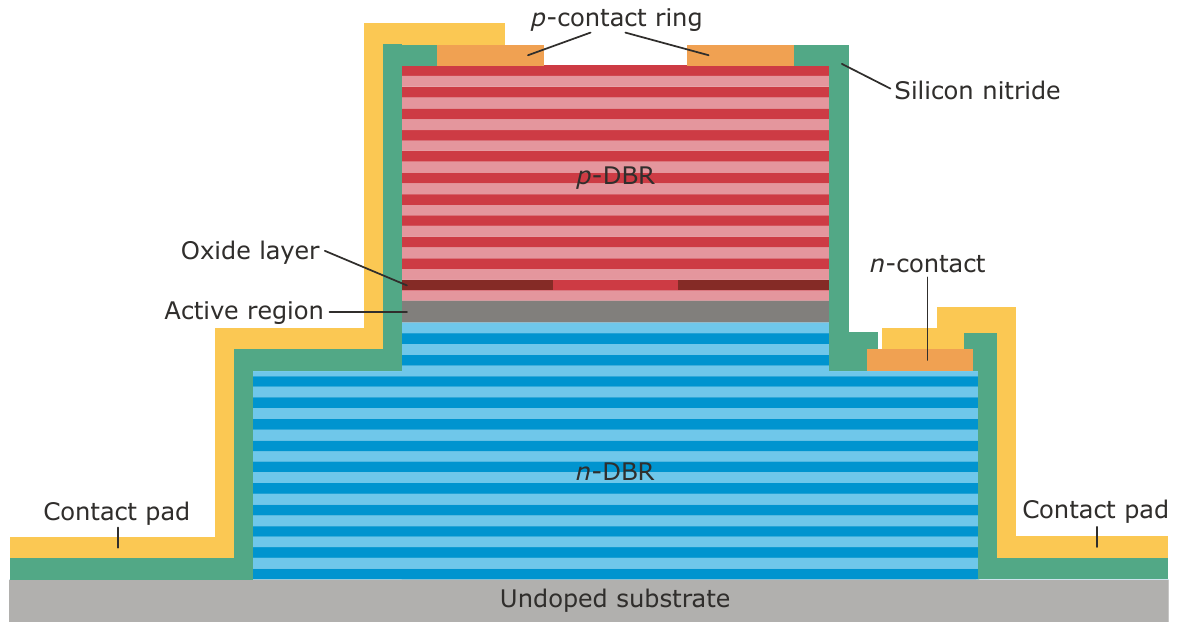}
\caption{Simplified VCSEL cross-section.}
\label{fig:vcselcross}
\vspace{-10pt}
\end{figure}
A simplified cross-section of a typical VCSEL is shown in Fig.~\ref{fig:vcselcross}. It contains an active region composed of quantum wells (QWs), and a separate confinement heterostructure (SCH) sandwiched between a top and a bottom distributed Bragg reflector (DBR) made of epitaxial (e.g., AlGaAs) or dielectric materials (e.g., SiO$_2$/TiO$_2$  \cite{Jahed_2019wavelength}). As the active region is thin (below 100 nm) and thus gain is limited, many pairs (on the order of 20-30) of DBR layers with width $\sim \lambda/4$ are required to create sufficiently high reflectivities ($>$99\%) \cite{Michalzik2013} to achieve lasing. Commonly the upper and lower DBRs are $p$- and $n$-doped and contacted via metal pads to enable electrical pumping of the active region. Transverse confinement of current and optical modes can be achieved using various methods. For datacom applications, oxidizing a high Al-content AlGaAs layer to provide a current isolating annulus \cite{doi:10.1063/1.113087}, called oxide aperture or oxide layer, is commonly used. This also provides an index step that guides the light inside the VCSEL.

The VCSEL differs from the edge-emitting laser (EEL) in many aspects. First and foremost, the geometry of the laser is different. The VCSEL has its cavity in the vertical direction, perpendicular to the substrate, and parallel to the current direction through the active area. In contrast, the EEL's cavity is formed parallel to the substrate by cleaving. Contrary to EELs, VCSELs have the distinct advantage of on-wafer testing during fabrication, lowering costs and maximizing device fabrication throughput. The circular geometry of the VCSEL provides yet another advantage over EELs, enabling an output beam that allows for easier fiber coupling. However, the VCSEL has a small volume and relatively high resistance and suffers from performance-lowering thermal effects.

The static and dynamic responses are of primary importance in studying the performance of a VCSEL and in the modeling of VCSELs. The dynamic response is further classified majorly into two essential subcategories: small-signal and large-signal response. 

\subsection{Static Figures of Merit}
\begin{figure}
\centering
\includegraphics[width=0.9\columnwidth]{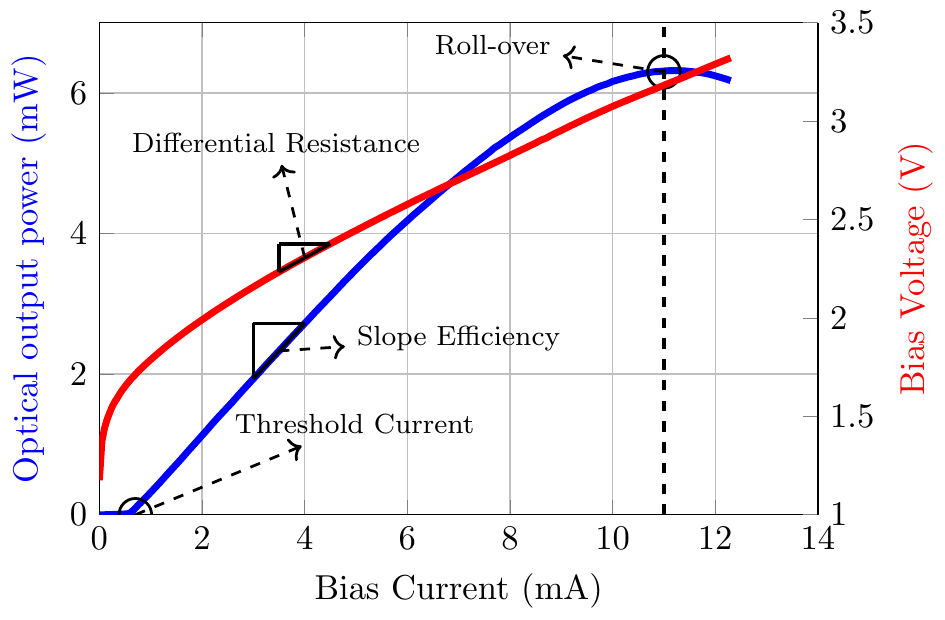}
\caption{Measured IPV of a VCSEL with marked figures of merit: threshold current, roll-over current, slope efficiency and differential resistance.}
\label{fig:AlexIPVFig}
\vspace{-10pt} \end{figure}

Static performance measurement is performed by sweeping the bias current $I_{b}$ of the VCSEL and then measuring the voltage applied to the VCSEL and the optical output power $P_{opt}$. Some static figures of merit (FoM) for a VCSEL for a fixed operating temperature can be seen in Fig.~\ref{fig:AlexIPVFig}, where a current-power-voltage (IPV)-measurement is shown. 
\begin{itemize}
    \item  \emph{Threshold current:} Threshold current is the current at which optical gain balances optical loss and the VCSEL "turns on". 

\item \emph{Slope efficiency:} It is measured in Watts per Ampere and relates the output power increase to the input current increase above the threshold. A low threshold current is associated with low resonator loss. However,  low-loss resonators out-couple little light, so output power and slope efficiency are low. Therefore, threshold current and slope efficiency are natural trade-offs. 

\item \emph{Roll-over:}
As expected for laser diodes, the light-current curve has a constant slope above the threshold but shows a roll-over for higher currents due to internal heating. The roll-over current signifies the current at which maximum optical output power is reached, after which thermal effects cause more heat to be generated and causes a drop in the output power.
\item \emph{Differential resistance:} It is the slope of the current-voltage curve and is measured in Ohms. It is non-linear and hence changes with the current. Low differential resistance is essential for lower heat generation.
\end{itemize}

Note that all the previously described operating characteristics will vary as a function of temperature, which is an important parameter to be included in the application design. 


\subsection{Dynamic Figures of Merit - Small-signal response}
The capacity of VCSEL-based OIs is directly related to the VCSEL dynamics. Data is encoded on the light emitted by the VCSEL by modulating the intensity through the input current. The VCSEL must be fast enough to react to changes in the current at the data-rate considered. This necessitates studying the dynamic response of the VCSEL, which can be described by a set of rate equations, which take into account the process behind and interactions between injected free carriers and photons in the cavity \cite{ColdrenBook}. The rate equations directly relate the active region's excess carrier density and the cavity's photon density with the current through the VCSEL. In other words, the rate equations relate the current with the output optical power $P_{opt}$. Perturbing the rate equations around a bias current $I_{b}$ using a first-order Taylor expansion and measuring the differential output power yields us the intrinsic small-signal modulation response, whose two-pole transfer function is \cite{ColdrenBook}
\begin{equation}
    H_{\text{int}}(f) = \eta_d \dfrac{hc}{\lambda_0 q}\cdot\dfrac{f^2_r}{f^2_r-f^2+j\gamma\dfrac{f}{2\pi}},
    \label{eq:transint}
\end{equation}
where $\eta_d$ is the differential quantum efficiency, $h$ the Planck constant, $c$ the speed of light, $\lambda_0$ the lasing wavelength in vacuum, $q$ the elementary charge, $f_r$ the resonance frequency and $\gamma$ the damping factor. The resonance frequency $f_r$ can be approximated as \cite{ColdrenBook}
\begin{equation}
    f_r \approx \dfrac{1}{2\pi}\sqrt{\dfrac{\upsilon_g g_0 S}{\tau_p\left(1+\varepsilon S\right)}},
    \label{eq:fr}
\end{equation}
where $\upsilon_g$ is the group velocity, $g_0$ the nominal differential gain $\text{d}G_0/\text{d}N$ ($G_0$ represents the unsaturated gain such that the material gain becomes $G = G_0/(1+\varepsilon S)$), $S$ the photon density in the cavity, $\tau_p$ the photon lifetime and $\varepsilon$ the gain compression factor. 

Summarizing equations \eqref{eq:transint} and \eqref{eq:fr}, a high differential gain, high photon density, and a short photon lifetime are all beneficial for increasing the VCSEL resonance frequency, which is desired, as this increases the modulation bandwidth of the VCSELs. 
VCSEL damping is also relevant for the modulation response --
severely under and overdamped VCSELs can constitute a problem for datacom applications. Underdamped VCSELs result in large over- and undershoot and ringing effects when modulated with data. Overdamped VCSELs, on the other hand, have too slow rise and fall times to be efficiently modulated. A flat (critically damped) or close-to-flat VCSEL response is the best for datacom applications.





\begin{figure}[]
\centering
\includegraphics[width=0.9\linewidth]{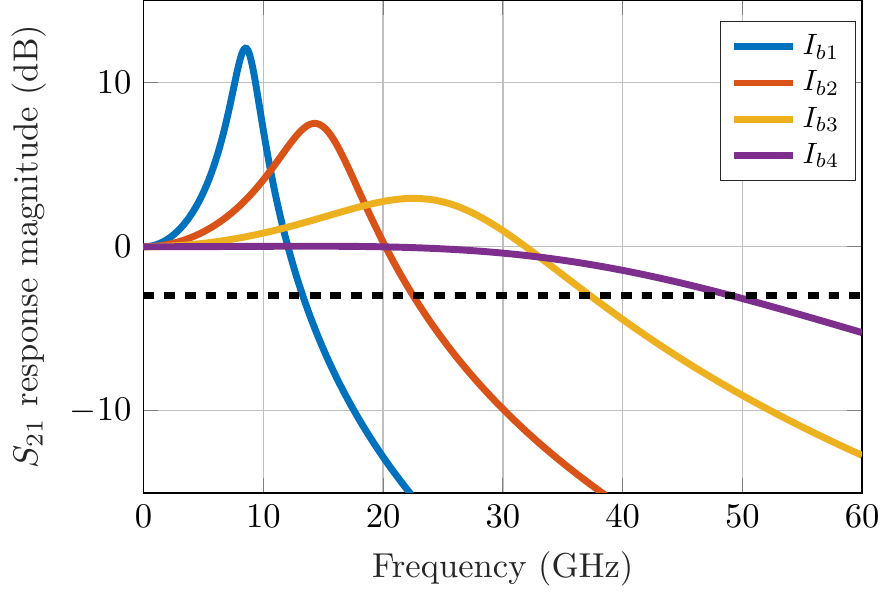}
\caption{Simulated intrinsic VCSEL response with parasitic effects neglected, for four representative bias currents $I_{b1}< I_{b2} < I_{b3} < I_{b4}$. The black dashed line indicates the $3~$dB bandwidth.}
\label{fig:AlexS21Fig}
\vspace{-10pt} \end{figure}

The small-signal modulation response is measured by $S_{21}$ 
\begin{equation}
    S_{21}= 20\log_{10}\frac{\abs{H_{\text{int}}(f)}}{\abs{H_{\text{int}}(0)}},
\end{equation}
and is plotted in Fig.~\ref{fig:AlexS21Fig} for increasing bias current, showing the movement of $f_r$ and that the VCSEL reaches a critically damped (flat) response at some current. 



\subsection{Dynamic Figures of Merit - Large-signal response}
The small-signal characteristics are insufficient to predict the entire performance of the VCSEL, as they do not capture the non-linear behavior of the device. A large-signal simulation is usually performed by back-to-back link simulation. For a modulation scheme of choice, a current source is used to bias the VCSEL, and a random bit sequence is used to supply a modulation voltage. They are combined to provide the drive current $I$ at the input of the VCSEL. The optical output power $P_{opt}$ is measured employing a photodetector. Eye diagrams are crucial to assess the large-signal FoMs of the VCSELs such as rise and fall time, overshoots,  deterministic jitter, and noise-margin.

VCSELs in datacom use IM/DD schemes exclusively. Typically, non-return-to-zero (NRZ) on-off keying (OOK) is used. However, at higher data-rates, a transition to PAM4 has occurred. So far, 8-level PAM (PAM8) has not been used in commercial datacom links. Instead, the data-rate and the number of lanes in OOK and PAM4 transmission have been increased to deal with higher bandwidth demand. Also, all parameters related to the modulation response must be considered and balanced for datacom VCSELs, with optimization for either OOK or PAM4, as the optimal VCSEL parameters are different for different modulation formats and baud rates \cite{Larsson2017vcselsdamping,Haglund2015Damping,Lengyel2017Damping}. 
\subsection{The Need for Modeling}


VCSEL modeling is an essential part of optimizing OIs because  the VCSEL is the primary bandwidth-limiting element in such links. While optical and O/E elements such as the fiber and photodetector can be modeled with low to moderate complexity with reasonable accuracy, higher-speed VCSELs require a more complex description due to their inherent nonlinearity. It is not a simple nonlinearity in which the instantaneous output is dependent just on the instantaneous input but on the entire non-zero span of the input. Furthermore, high-speed VCSELs in datacom require optimization of different parameters than, e.g., VCSELs for sensing. Since the intrinsic bandwidth of VCSELs is high, minimizing limitations imposed by thermal and parasitic effects is vital especially in co-packaged optics.\footnote{From a structural perspective, there are two characteristics that delineate a high-speed VCSEL. A high-speed VCSEL typically employs short active regions
designed with strained InGaAs QWs inside an AlGaAs SCH to increase the differential gain \cite{Healy_2010}. Thermal management is mainly done by modulation doping and interface grading of the DBRs, where heat is mainly generated through resistive Joule heating and free carrier absorption (FCA).} 

In short, it is imperative to develop models that can accurately capture for various temperatures, 
\begin{enumerate}
    \item the static response,
    \item the small-signal response or the modulation response, 
    \item and the large-signal response evaluated by means of the eye diagram.
\end{enumerate}In the following section, we discuss in detail the several VCSEL-modeling approaches available.


\section{Surrogate models for VCSELs}\label{sec:modeling}

 We begin this section by reviewing the classic circuit models and providing one such circuit-model example. We then motivate the necessity of developing suitable models for use in an E2E-learning setup. In this context, we introduce Volterra-series-based models and NN models. We finally provide some discussions on the opportunities in ML-based modeling.  

\subsection{Circuit Models}

The main large-signal VCSEL modeling efforts have focused on VCSEL circuit-level models, popularized in the 90s by P.V. Mena \cite{Mena1,Mena2}. This approach enables efficient and fast co-simulation with driving electronics and can use a physical rate equation-based description of the VCSEL combined with essential input impedance circuit elements. Circuit-level VCSEL models, such as \cite{Li1,Liang1,Bruensteiner1999,Yan2019,grabowski2021large,Belfiore2016,Mena1,Wang2016,Zhang2021,szczerba2018behavioral,Mena2}, are often hybrid models that employ rate equations, and therefore fall in-between a purely physical and empirical description. While physical models might provide critical feedback to device designers, it is harder to fit simulations to measurements and extract physical parameters. On the other hand, empirical models are easier to fit and can reproduce the behavior better. However, they provide less insight into the physics of the device. 

The essential parts of any circuit-level VCSEL model include an input impedance sub-circuit, tracking of the carriers in the QWs and photons in the resonator, and thermal effects.

\begin{figure}
\centering
\includegraphics[width=\linewidth]{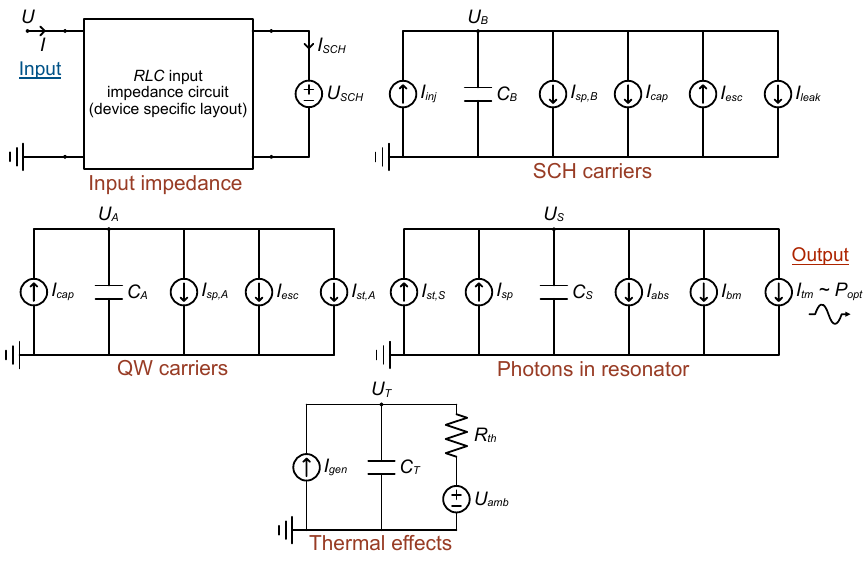}
\caption{A circuit-level large-signal equivalent model for datacom VCSELs, with five interdependent sub-circuits each governing a group of physical phenomena, For a detailed discussion on what each term signifies, refer \cite{grabowski2021large}.}
\label{fig:AlexModelFig}
\vspace{-10pt} \end{figure}

\begin{figure}
\centering
\includegraphics[width=0.9\linewidth]{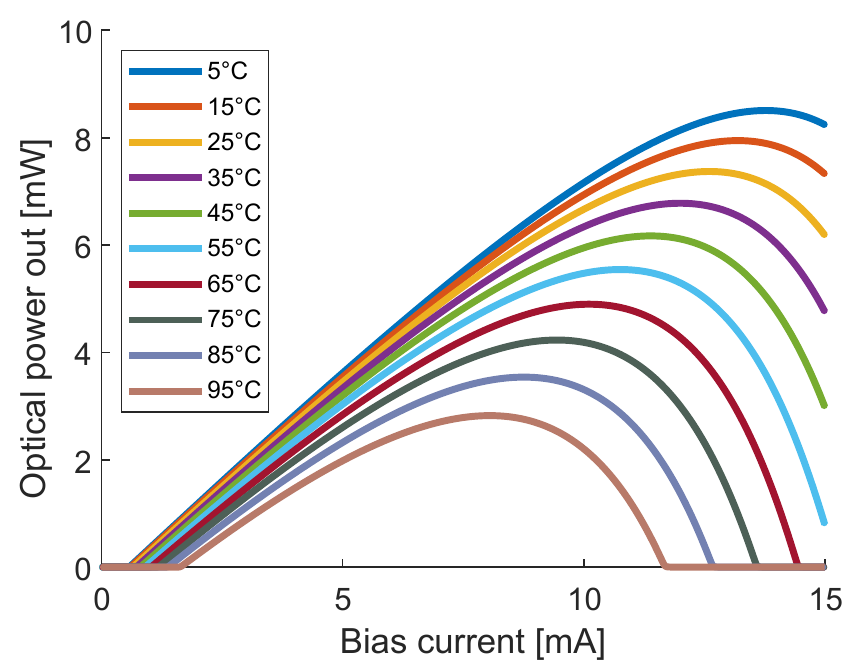}
\caption{Simulated optical power from the VCSEL vs bias current, for different ambient temperatures.}
\label{fig:AlexPoptFig}
\vspace{-10pt} \end{figure}
An example of a circuit model can be seen in Fig.~\ref{fig:AlexModelFig}, where five interdependent sub-circuits are shown, and each keeps track of a group of physical phenomena: the input impedance, the carriers in the QWs, the carriers in the SCH, the photons in the resonator and thermal effects:
\begin{enumerate}
    \item The input impedance is typically modeled by including the VCSEL pad, mesa, and active area. 
    Impedance modeling of the active layer typically includes either only RC-elements \cite{Li1,Liang1,Yan2019} 
    or RC-elements together with voltage sources \cite{grabowski2021large,Belfiore2016,Mena1,
    szczerba2018behavioral}, often in combination with diodes. \emph{At the input of this circuit, the current $I$ and the voltage $U$ represent the VCSEL drive current (biased around $I_b$) and VCSEL voltage drop, respectively.}
    \item 
    The carrier tracking can be separated into two sub-circuits, one for the more essential QW-bound carriers, which dictates the stimulated and spontaneous emission rates, and the other for the SCH continuum state carrier tracking. 
    \item The photons created in the resonator through stimulated and spontaneous emission must be accurately tracked, as the rates of mirror loss and free carrier absorption depend on the photon density. This also dictates the output power through the VCSEL surface. \emph{At the output of this circuit, the current $I_{\text{tm}}$ provides the optical output power from the VCSEL, $P_{\text{opt}}$ such that $P_{\text{opt}} \propto I_{\text{tm}}$}.
    \item  Thermal effects are usually modeled using a thermal impedance description, with an offset for the ambient temperature and a thermal time constant to predict the dynamic thermal behavior. 
    \item  Apart from the sub-circuits mentioned above, noise should be included in VCSEL models. Since phase noise is of no interest in IM/DD links, only intensity noise should be included, either through an amplified spontaneous emission approximation or on a per-process basis.
\end{enumerate}
Such circuits can reproduce VCSEL static behavior, e.g., optical output power vs. input current shown in Fig.~\ref{fig:AlexPoptFig}, and also the dynamic behavior in the small-signal domains (the modulation response) and the large-signal domains (the output power eye-diagrams) accurately. However, the circuit-level implementations suffer from two drawbacks: convergence of simulations and the requirement of complex full-link simulations. In other words, since they are restricted to use with circuit simulators, the entire link must be modeled in a circuit simulator. Furthermore, circuit-level models are more critical to IC design but are not amenable to E2E learning using AEs. Also, the differential equations governing the VCSEL may also not be directly amenable to gradient back-propagation and depend on the specifics of the employed differential equations solver \cite{chen2018neural}. Therefore, incorporating the rate equations in the E2E optimization of AEs is challenging. It is necessary to develop "ML-friendly" models that capture the operating characteristics of VCSELs.

\subsection{Volterra Series}
\begin{figure}[]
\centering
\begin{subfigure}[t]{0.48\linewidth}
\includegraphics[width=\textwidth]{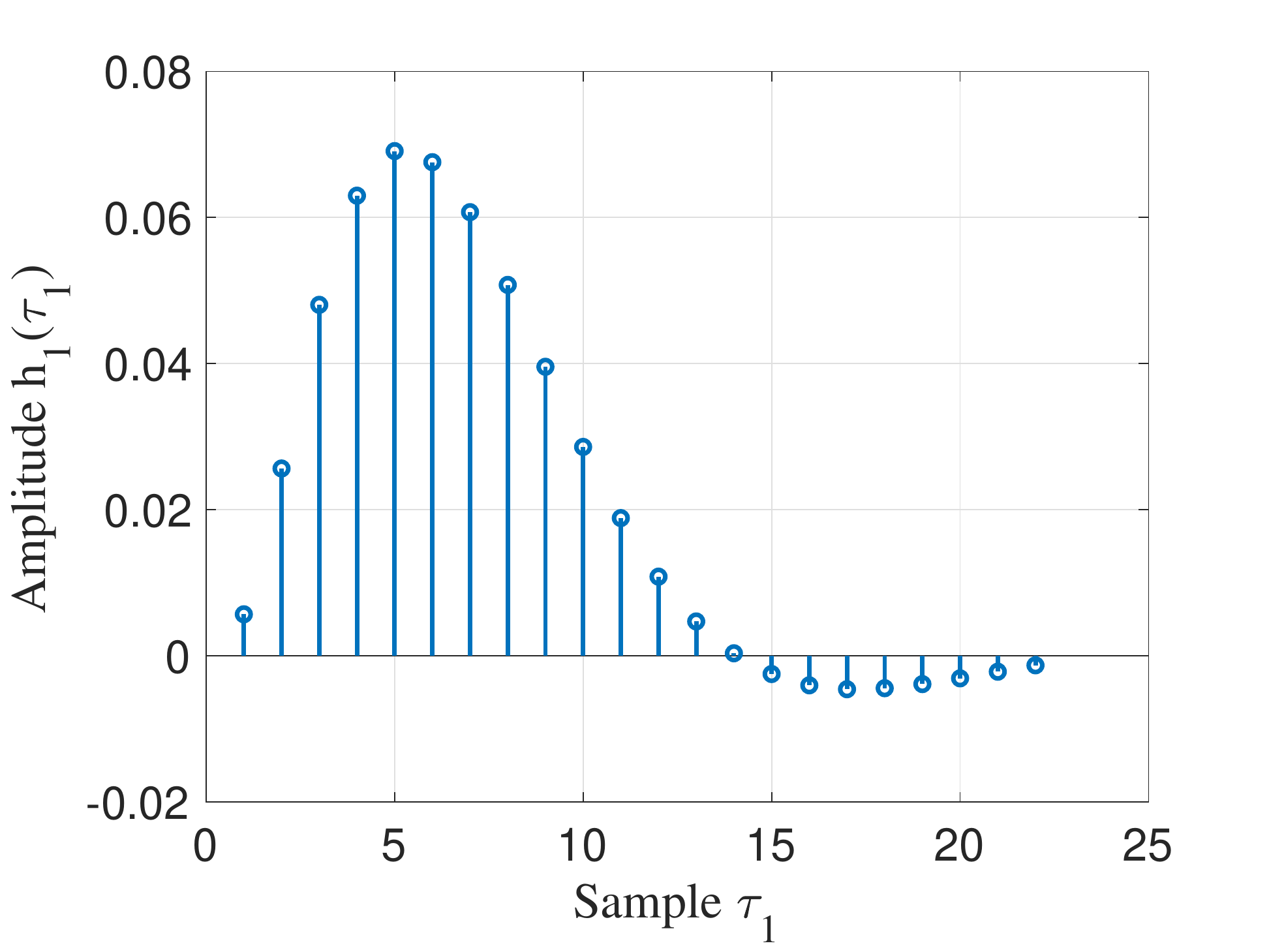}
\caption{Amplitude of the first order kernel - $h_1(\tau_1)$- describes the impulse response.}
\label{fig:volt_1st}
\end{subfigure}%
\hfill
\begin{subfigure}[t]{0.48\linewidth}
\includegraphics[width=\textwidth]{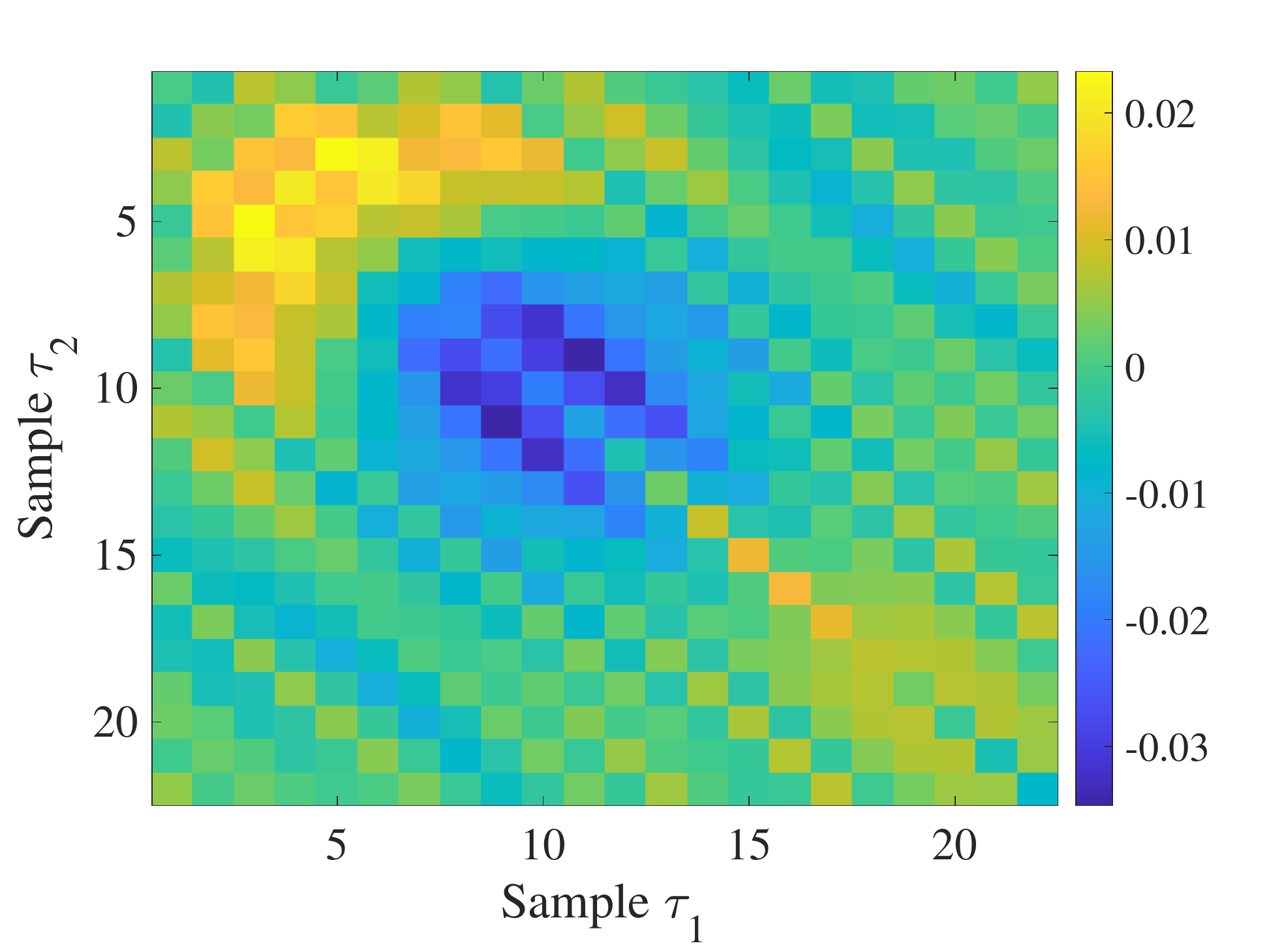}
\caption{Amplitude of the second order kernels - $h_2(\tau_1, \tau_2)$- shown as a 2D grid plot - describes the nonlinearity.}
\label{fig:volt_2nd}
\end{subfigure}
\caption{Coefficients of the Volterra kernels.}
\vspace{-10pt} \end{figure}

\begin{figure}[]
\centering
\begin{subfigure}[t]{0.48\linewidth}
\includegraphics[width = \textwidth]{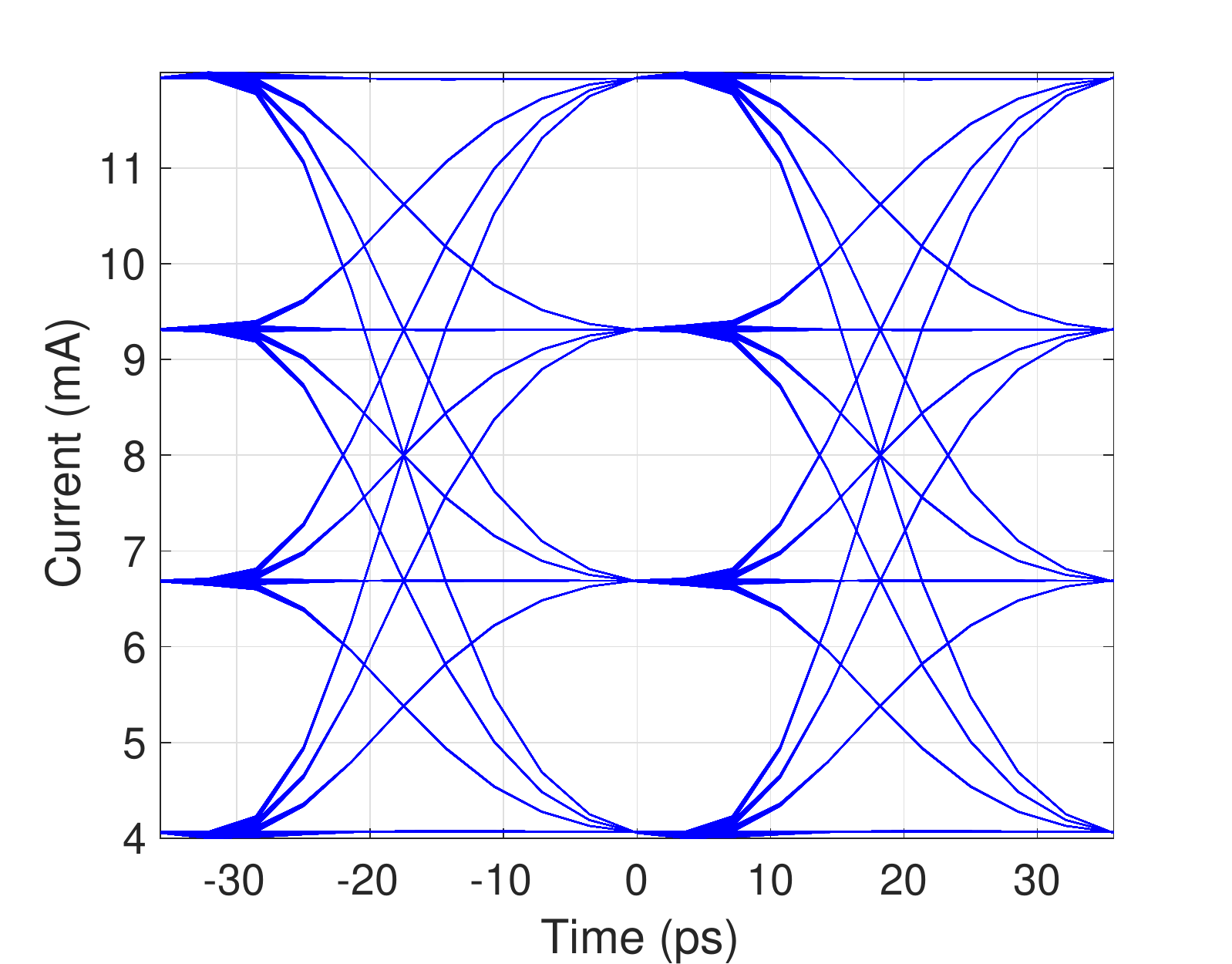}
\caption{Drive current.}
\label{fig:volt_stim_eye}
\end{subfigure}%
\begin{subfigure}[t]{0.48\linewidth}
\includegraphics[width = \textwidth]{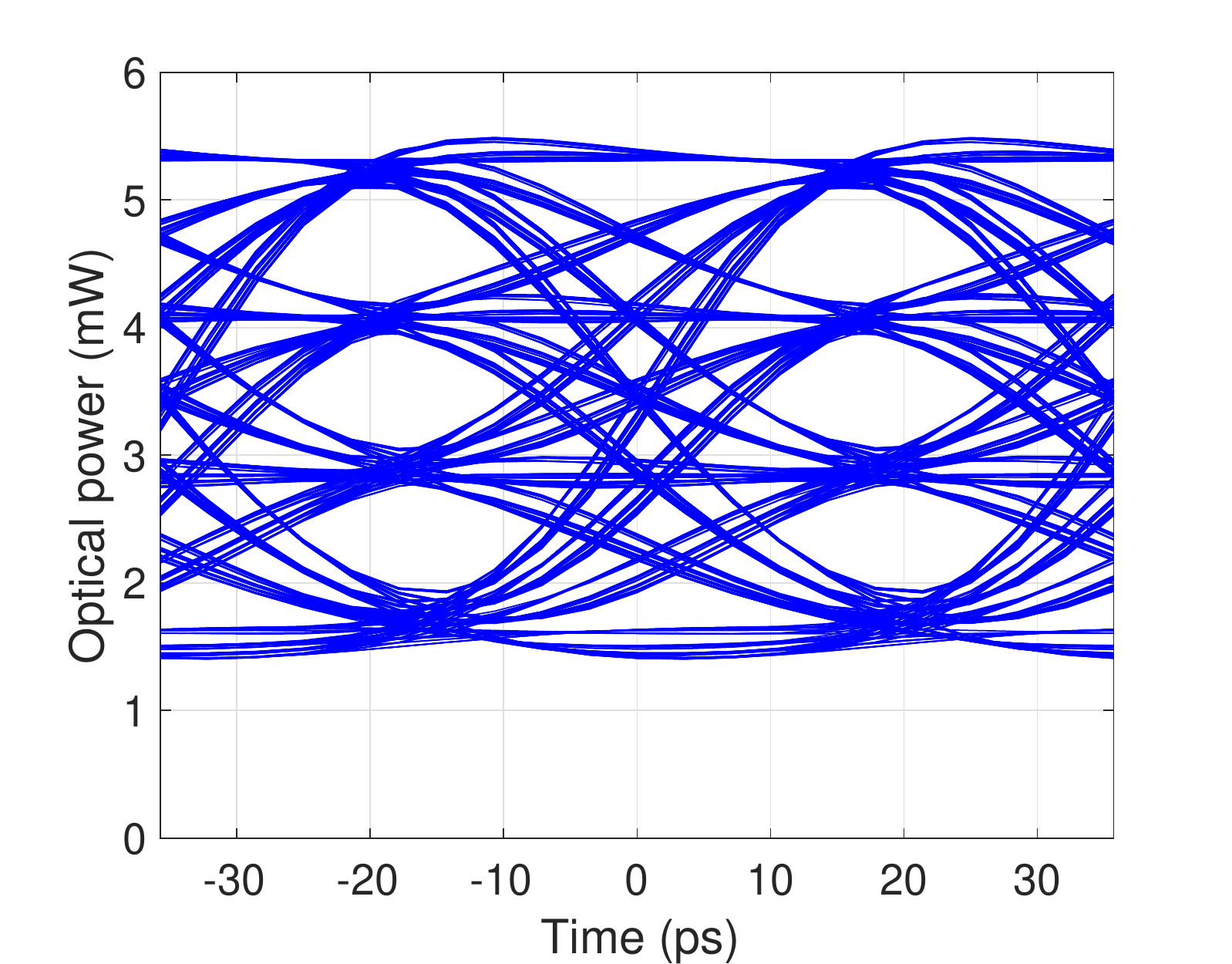}
\caption{Rate equation model.}\vspace{3mm}
\label{fig:volt_re_eye}
\end{subfigure}
\begin{subfigure}[t]{0.48\linewidth}
\includegraphics[width = \textwidth]{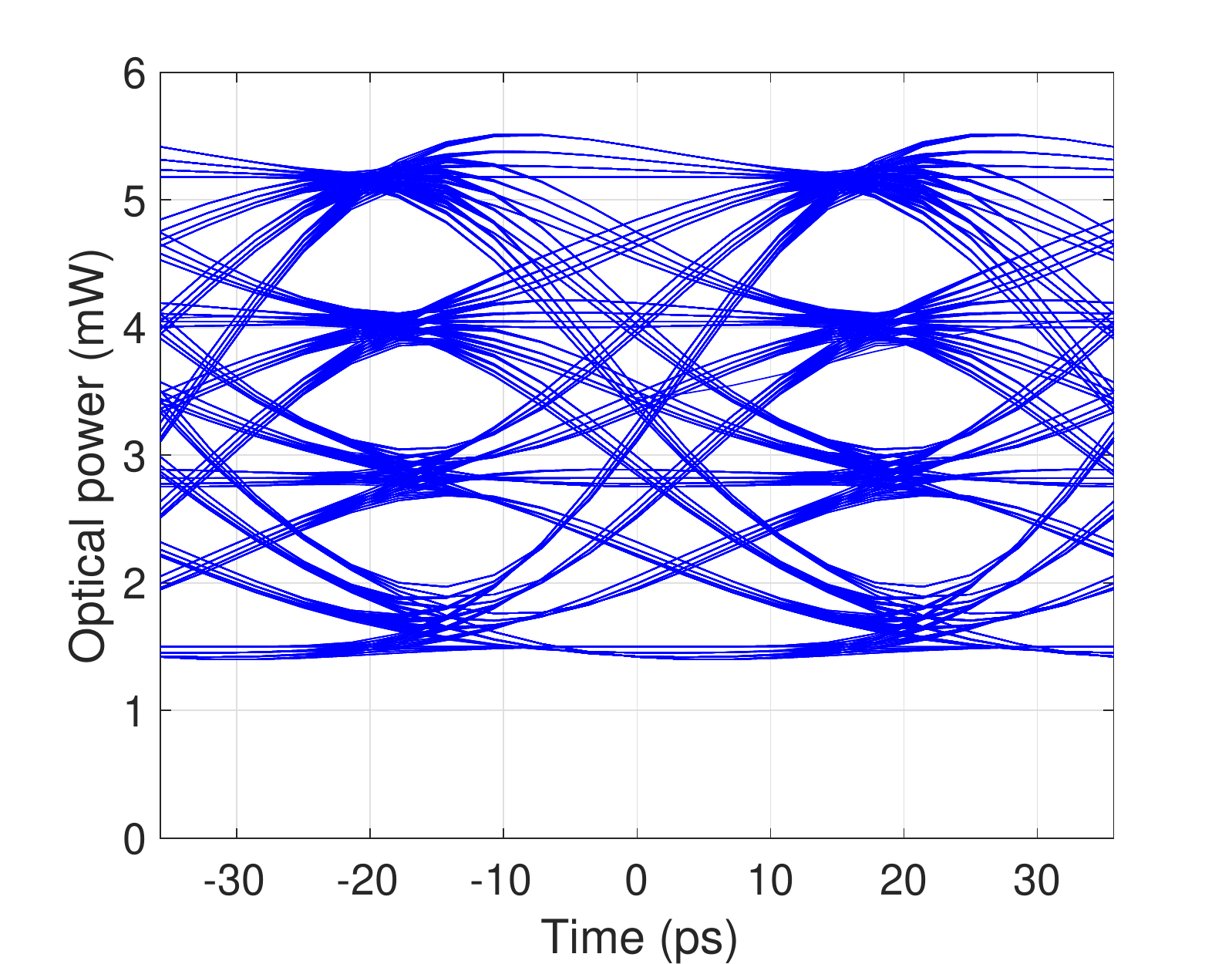}
\caption{Volterra series.}
\label{fig:volt_volt_eye}
\end{subfigure}%
\begin{subfigure}[t]{0.48\linewidth}
\includegraphics[width = \textwidth]{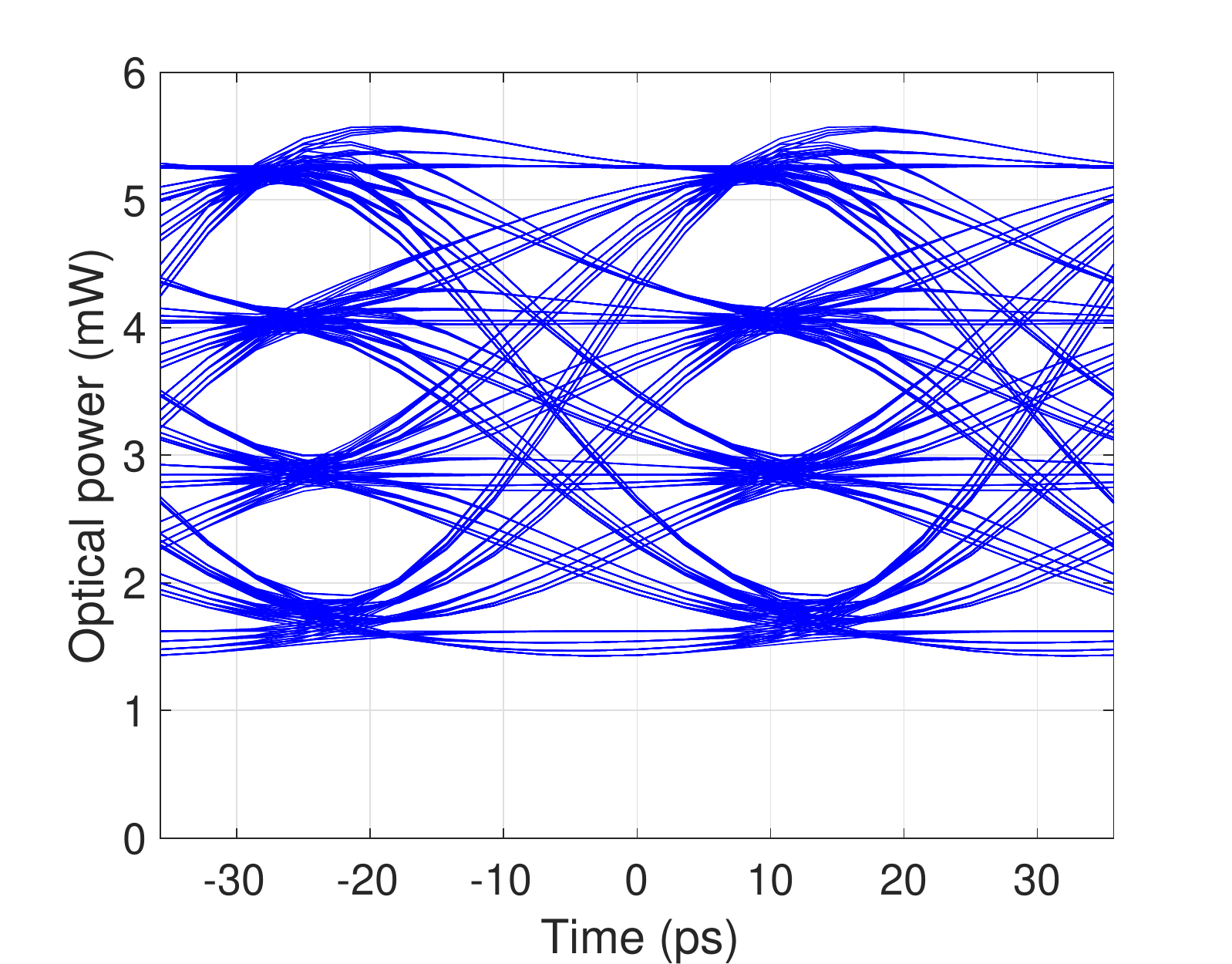}
\caption{TDNN model.}
\label{fig:tdnn_eye}
\end{subfigure}
\caption{56 Gbps PAM4 eye diagram of the drive current (a), rate equation (b), Volterra series (c) and TDNN (d) VCSEL models.}
\vspace{-10pt} \end{figure}

A tool that enables the modeling of the nonlinear behavior of the VCSELs and can easily fit into an E2E AE is the Volterra series. 
A discrete system with input $x(n)$ and output $y(n)$ for $n \in \mathbb{Z}$ can be expanded into a $P$-th order Volterra series as
\begin{equation}
    \label{eq:volt_gen}
    \begin{split}
    y(n) & = h_0 + \\ 
    & \displaystyle\sum_{p=1}^{P}\displaystyle\sum_{\tau_1=a}^{b}\cdots\displaystyle\sum_{\tau_p=a}^{b}h_p(\tau_1,...,\tau_p)\displaystyle\prod_{j=1}^{P}x(n-\tau_j),
    \end{split}
\end{equation}
where $h_0$ is a constant, $h_p(\tau_1,...,\tau_p)$ are the $p$-th order discrete-time Volterra kernels and $\tau_1, \hdots \tau_p$ are time indices. When limited to second order, (\ref{eq:volt_gen}) reduces to
\begin{equation}
    \label{eq:volt2nd}
    \begin{split}
    y(n) & = h_0 + \displaystyle\sum_{\tau_1=a}^{b}h_1(\tau_1)x(n-\tau_1) + \\
    & \displaystyle\sum_{\tau_1=a}^{b}\displaystyle\sum_{\tau_2=a}^{b}h_2(\tau_1,\tau_2)x(n-\tau_1)x(n-\tau_2).
    \end{split}
\end{equation}
The second-order Volterra series, as given by (\ref{eq:volt2nd}), has three components in the sum: the DC offset component, the convolution of the input signal with the first-order kernel, which is an impulse response, and finally a two-dimensional convolution term with a two-dimensional kernel, which describes the scaling of a product of time-delayed versions of the input signal. 

\begin{example}
We present an example of applying a second-order Volterra series to a VCSEL model, with the input drive current $I$ represented by the input $x(n)$ and the optical output power $P_{opt}$ represented by the output $y(n)$. A simple rate equation numerical model was first implemented to obtain the white Gaussian training data source of $6~$mA standard deviation at $8~$mA bias current. The Volterra kernels were then estimated using the Lee--Schetzen’s correlation method \cite{orcioni2014improving}.
Identified first-order kernels are shown in Fig.~\ref{fig:volt_1st}, and the second-order kernels are shown in Fig.~\ref{fig:volt_2nd}. 
To compare the Volterra series with the baseline rate equation model, $56$ Gbps PAM-4 data was set as a stimulus to both models. 
The eye diagram of the model stimulus signal is shown in Fig.~\ref{fig:volt_stim_eye}, the rate equation model output is illustrated in Fig~\ref{fig:volt_re_eye}, and the Volterra model output is illustrated in Fig.~\ref{fig:volt_volt_eye}. The normalized root-mean-square error (NRMSE) between the Volterra and the rate equation models was $2\%$. The coefficient of determination\footnote{The coefficient of determination is a number between 0 and 1 that measures how well a statistical model predicts an outcome.} was $0.9959$. The Volterra series is inherently differentiable, accurately models the large-signal response, and, therefore, can be used as a surrogate model for the VCSELs in AE-based E2E learning.
\end{example}

\subsection{Neural Network (NN) based VCSEL models}
Just as the nonlinear transfer function was modeled using the Volterra series, it can be modeled using a NN. NNs are information processing systems comprising an input layer, one or more hidden layers---where intermediate processing or computation is done---and an output layer containing multiple nodes. The nodes in two adjacent layers are interconnected with an assigned variable weight. The activation function, which models the nonlinearity, defines the output of a node given an input. The forward-propagation process provides the inference given input, and the backpropagation updates the weights of the nodes until the chosen loss function, which compares the target and predicted output values, is minimized. 

\begin{figure}[t!]
    \centering
            \includegraphics[scale=0.80]{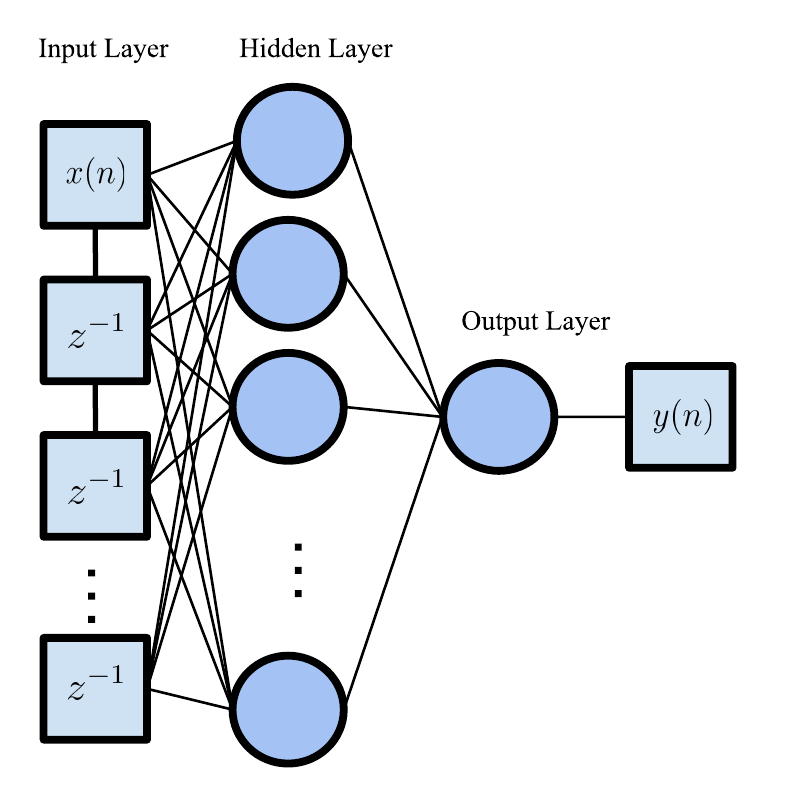}
            \caption{A block diagram of a TDNN with one hidden layer.}
            \label{fig:tdnn}
\vspace{-10pt} \end{figure}

Since the VCSEL drive current and the optical output power are time functions, an appropriate NN structure is the time domain neural network (TDNN), traditionally used in speech recognition \cite{waibel1989modular}. TDNNs are a subset of convolutional neural networks (CNNs) working on one-dimensional input. The most basic TDNN to model a VCSEL has an input layer with some delays to hold the input waveform samples, a single hidden layer, and an output layer, as shown in Fig.~\ref{fig:tdnn}. 

\begin{example}In our case, to model the VCSELs, the input layer has 22 delays, and the hidden layer has 22 elements and uses a hyperbolic tangent sigmoid activation function. The network is trained using the same type of signal which was used to identify the Volterra kernels, i.e., white Gaussian noise applied to the rate equation model. The NN parameters are updated during the backpropagation using a variation of stochastic gradient descent, and the loss function is the mean-square error (MSE). The large-signal output to the $56$ Gbps PAM-4 input drive current is shown in Fig.~\ref{fig:tdnn_eye}. The resultant eye diagram is very close to the Volterra series and the rate equation model. The coefficient of determination was $0.9991$, and the NRMSE was $1\%$, which is better than the Volterra series. 
\end{example}

\subsection{Opportunities in ML-based modeling of VCSELs}
Both the Volterra and the TDNN models are differentiable and can be used as VCSEL surrogates to train an AE. They directly relate the optical output power to the input drive current. They can potentially capture the large-signal response of the VCSELs with a high degree of precision. However, there is no guarantee that such a NN would capture all the characteristics of VCSEL, especially the bias current vs. optical output power (Fig. \ref{fig:AlexPoptFig}) or the small-signal response (Fig. \ref{fig:AlexS21Fig}) including the thermal effects, across all operating temperatures, with limited training samples.\footnote{An increase in training samples may also lead to over-fitting and incorrect results.} A NN model that captures the small-signal response can be applied to varying operating data-rates without retraining them at each required data-rate. Furthermore, capturing the thermal effects, such as the roll-over, would help optimize the PAM-4 levels accurately in an E2E learning setup. Therefore, a surrogate NN model that captures the VCSEL dynamics in its entirety calls for the exploitation of other learning techniques.  

Recently, physics-informed neural networks (PINNs) were proposed in \cite{raissi2019physics}, which augment NN training by incorporating physical laws in the form of differential equations governing an underlying data set. PINNs have been used and explored in several fiber-optics applications\cite{wang2022applications, jiang2022physics}. 
Such computationally efficient surrogate models infer solutions to general nonlinear partial differential equations utilizing the multiple layers of neural units and employing a loss function satisfying the differential equations to be solved. Since differential equations govern the operations of VCSELs, a PINN-inspired surrogate model can be explored.

\section{Component-specific Learning and Compensation}\label{sec:Component}

\begin{figure}[t!]
    \centering
            \includegraphics[scale=0.95]{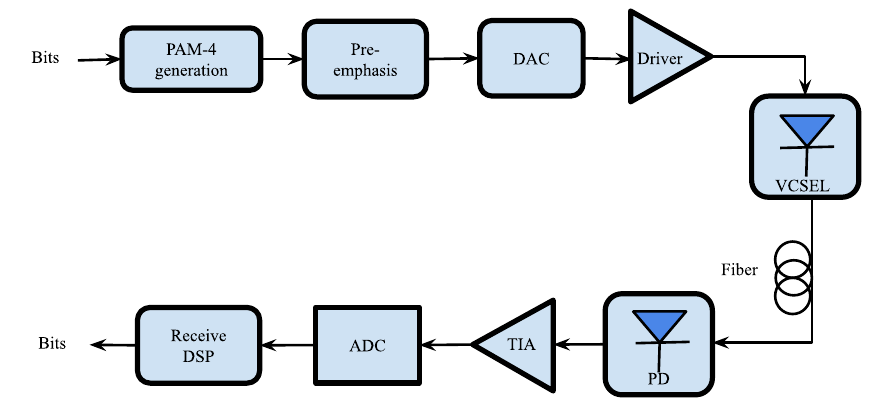}
            \caption{A model block diagram of a VCSEL-based OI system. Here DAC represents digital-to-analog converter, ADC represents analog-to-digital converter, PD represents photodetector and TIA represents trans-impedance amplifier.}
            \label{fig:E2E-opticalinterconnects}
\vspace{-10pt} \end{figure}

Fiber-optic communication is affected by nonlinearities due to the physical properties of the fiber, including the Kerr effect. However, the fiber is not the only source of nonlinear distortion. Many O/E components, including VCSELs, contribute to nonlinear transfer characteristics in the optical transceiver architecture. Fig.~\ref{fig:E2E-opticalinterconnects} shows a general schematic of a VCSEL-driven OI system, highlighting the parts of the transmission link where digital-to-analog/analog-to-digital, electro-optical/opto-electrical conversions, and optical fiber propagation are performed. To truly achieve robust 100 Gbps links, the judicious use of equalization techniques to compensate for the nonlinear effects of the VCSELs becomes imperative. Equalization---either post-equalization at the receiver or pre-equalization at the transmitter---is vital for compensating the impairments caused by the limited bandwidth of O/E components such as VCSELs. In this section, we will first review the use of NNs as a post-equalizer and then as a pre-equalizer or pre-distortion compensator.
\subsection{Post Equalization}\label{sec:post}
A variety of different digital signal processing (DSP) based equalizers are commonly employed to reduce/cancel inter-symbol interference (ISI) in VCSEL-based OIs, including feed-forward equalizer (FFE)\cite{kuchta2016higher}, decision feedback equalizer (DFE), maximum likelihood sequence estimator (MLSE) \cite{karinou2015experimental, tan2017high}, and Volterra series-based equalization (VSE) \cite{liu2017high}. 
However, many nonlinearities in a practical system can only be approximately captured by such models and are challenging to equalize by conventional model-based DSP methods. 
Therefore, ML-based DSP algorithms have been gaining popularity, such as support vector machine nonlinear equalizer \cite{giacoumidis2017reduction, liang2018experimental} or deep belief network-hidden Markov model-based equalizers \cite{tian2020deep}. 

The use of NNs for equalization has also gained popularity in recent times. The most popular NN structure is the fully connected neural network (FCNN). FCNNs are employed in many works as an equalizer and have been shown to provide excellent nonlinear equalization performance \cite{gaiarin2016high,chuang2017employing,  xu2019computational, liao2020unsupervised}. In \cite{gaiarin2016high}, the equalization performance of an FCNN is observed to be nearly equal to that of a 21-tap FFE. In \cite{chuang2017employing}, a FCNN-based equalizer is shown to outperform the $3$rd-order VSE with significantly less mathematical complexity. Furthermore, \cite{xu2019computational} demonstrates that only tens of multiplications are needed for the studied NN-based equalizers to achieve bit error rate (BER) performances below the hard-decision (FEC) threshold. In \cite{liao2020unsupervised}, it is demonstrated that an unsupervised learning scheme for FCNN equalizers in IM/DD system can achieve the same performance as the model trained by the traditional supervised learning method. 

Besides FCNNs, CNNs \cite{chuang2018convolutional}, recurrent neural networks (RNNs)\cite{lavania2015adaptive, zhou2019low, qin2019low, xu2019computational, xu2020feedforward}, and long short-term memory (LSTM) NNs~\cite{oh2022bi} are also used to improve the equalization performance at the cost of computational complexity and training epochs. However, it is possible to use transfer learning to use the knowledge gained from NNs trained for equalization with FCNN and train an RNN with a reduced number of epochs \cite{xu2020feedforward}. 
However, the practical deployment of real-time NN-based channel equalizers requires that their computational complexity is, at least, comparable or desirably lower than that of existing conventional DSP solutions. In order to decrease the complexity and the corresponding hardware requirements, iterative pruning and quantization techniques have been investigated in \cite{ge2020compressed,sang2022low}.

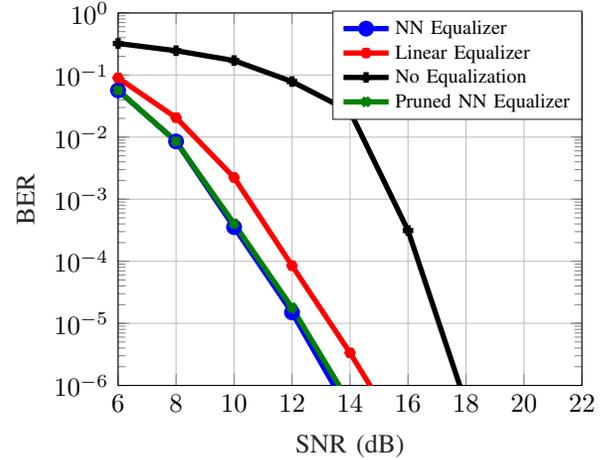
\begin{figure}[t]
\centering
\definecolor{mycolor1}{rgb}{0.00000,0.49804,0.00000}%
\begin{tikzpicture}

\begin{axis}[%
width=2.428in,
height=1.951in,
at={(1.011in,0.644in)},
scale only axis,
xmin=6,
xmax=22,
xlabel style={font=\color{white!15!black}},
xlabel={SNR (dB)},
ymode=log,
ymin=1e-06,
ymax=1,
yminorticks=true,
ylabel style={font=\color{white!15!black}},
ylabel={BER},
axis background/.style={fill=white},
xmajorgrids,
ymajorgrids,
legend style={at={(0.462,0.713)}, nodes={scale=0.75, transform shape}, anchor=south west, legend cell align=left, align=left, draw=white!15!black}
]
\addplot [color=blue, line width=2.0pt, mark=o, mark options={solid, blue}]
  table[row sep=crcr]{%
6	0.056705103459311\\
8	0.008510765968937\\
10	0.000355031952876\\
12	1.5001350122e-05\\
14	3.5401356761e-07\\
16	0\\
18	0\\
20	0\\
};
\addlegendentry{NN Equalizer}

\addplot [color=red, line width=2.0pt, mark=asterisk, mark options={solid, red}]
  table[row sep=crcr]{%
6	0.090078107029633\\
8	0.020581852366713\\
10	0.002235201168105\\
12	8.5007650689e-05\\
14	3.3550456770e-06\\
16	1e-07\\
18	0\\
20	0\\
};
\addlegendentry{Linear Equalizer}

\addplot [color=black, line width=2.0pt, mark=+, mark options={solid, black}]
  table[row sep=crcr]{%
6	0.322570458455015\\
8	0.246576964235708\\
10	0.171230172620714\\
12	0.0784675411049326\\
14	0.0259747419690363\\
16	0.000313037564507741\\
18	5.00060007200864e-07\\
20	0\\
};
\addlegendentry{No Equalization}

\addplot [color=mycolor1, line width=2.0pt, mark=x, mark options={solid, mycolor1}]
  table[row sep=crcr]{%
6	0.056705103459311\\
8	0.008510765968937\\
10	0.000405031952876\\
12	1.8001350122e-05\\
14	5.1073452312e-07\\
16	0\\
18	0\\
20	0\\
};
\addlegendentry{Pruned NN Equalizer}

\end{axis}

\end{tikzpicture}%
\caption{BER vs SNR for several equalizers are shown. A nonlinear NN equalizer provides BER performance when compared to the case of absence of any equalization.}
\label{fig:bervssnr}
\vspace{-10pt} \end{figure}

\begin{example} (NN-based equalizers) We demonstrate the BER performance of a post-equalizer NN that uses FCNN with the time series signal fed as a 1-D input array similar to that in \cite{chuang2017employing}. The input array comprises samples of past symbols to account for the system's inter-symbol interference (ISI). The hidden layer has 4 neurons with ReLU activation function. The output layer has $4$ neurons corresponding to $4$ levels of the assumed PAM constellation. The labels of the symbols are one-hot vectors, and the loss function is the cross-entropy \footnote{We note that a criteria like block error rate cannot be used here, since they do not allow gradient backpropagation.}. Fig.~\ref{fig:bervssnr} shows the bit error rate (BER) performance of the 2-layer NN-based equalizer for a PAM-4 modulated 2-symbol long (oversampled by 10) input time series fed through the rate equation numerical model of VCSEL. The NN equalizer provides nearly a $4~$dB sensitivity gain over the case with no equalization for a BER of $10^{-6}$ and a $1~$dB sensitivity gain over a linear FFE also implemented employing a NN with linear activation function. 

To demonstrate pruning the equalizer, after initial training, we used the Tensorflow API called Tensorflow Model Optimization that eliminates the smallest weights at the end of every training step following a polynomial decay schedule \cite{abadi2016tensorflow}. At the end of $500$ such training steps, we achieved a final weight sparsity of $60\%$. This translates to a drop in the number of multiplications to nearly $20$ when compared to $100$ for the unpruned network. Also, from Fig.~\ref{fig:bervssnr}, we can observe that such a pruned NN equalizer performs nearly as well as a non-pruned NN equalizer.  
\end{example}

\subsection{Nonlinearity Pre-compensation}\label{sec:pre}

Digital equalization is also commonly used at the transmitter side, referred to as digital pre-compensation. 
The main idea is to pre-distort the transmitted waveforms in the digital domain prior to transmission to improve the E2E performance or to reduce the receiver complexity \cite{berenguer2015nonlinear,lavery2017bandwidth}. 
To date, many digital pre-compensation efforts have focused on transmitter hardware impairments compensation, known as digital pre-distortion (DPD).\footnote{We note that digital pre-compensation can also be used for fiber nonlinearity compensation (e.g., see \cite{lavery2017bandwidth}). For a VCSEL-based system, the fiber nonlinearity is negligible due to the low VCSEL output power and short communication reach. Therefore, we leave out the discussion about digital pre-compensation for fiber nonlinearity mitigation.} Conventionally, linear static\cite{DACcomp1} or dynamic\cite{DACcomp2,DACcomp3} finite impulse response (FIR) filters have been widely applied to compensate for the bandwidth restrictions. Nonlinear schemes such as those using the $\arcsin$ function combined with clipping can effectively reduce the performance loss originating from the Mach-Zehnder modulator nonlinearity and digital-to-analog converter quantization noise\cite{Tximpair}. However, linear DPD combined with the inverse modulator response cannot entirely mitigate the transmitter distortions and more robust approaches are needed to cope with imperfections of current high-rate optical communication systems\cite{pilotDSP}.

Over the last decade, more complex algorithms using the Volterra series (or its variants such as memory polynomials) have been studied extensively\cite{Volterra1, Volterra2, MP}, showing promising performance at the expense of significantly increased complexity. 
As an alternative, DPD based on NNs has attracted much interest in recent years, showing that NN-based DPD can achieve better/similar performance as those using Volterra series, but with reduced complexity using LSTM \cite{paryanti2020direct} and TDNN \cite{wu2022low}.  

\begin{figure}[t]
    \centering
    \includegraphics[width=1\columnwidth]{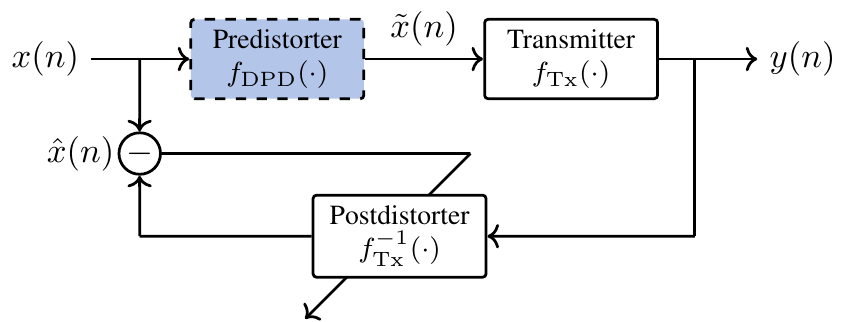}
    \caption{Block diagram of the ILA. Note that in the post-distorter training phase, the pre-distorter is not in use (i.e., $\tilde{x}(n)= x(n)$). The post-distorter is learned by minimizing the MSE between its output $\hat{x}(n)$ and the transmitted waveform $x(n)$. The learned post-distorter is then used as the pre-distorter.}
    \label{fig:ILA}
\vspace{-10pt} \end{figure}

The goal of DPD, denoted by $f_\mathrm{DPD}(\cdot)$, is to pre-distort the transmitted waveform $x(n)$ such that the difference (i.e., the mean squared error (MSE)) between $x(n)$ and the transmitter output $y(n)$ is minimized, which we write as

\begin{align}
     \label{DLA-objective}
    \hat{f}_{\mathrm{DPD}} = \underset{{f_\mathrm{DPD}}}{\arg \min} \frac{1}{N}\sum_{n=1}^{N} |y(n)- x(n)|^2,
\end{align}
where $y(n)=f_\theta(f_\mathrm{DPD}(\mathbf{x^{(n)}}))$, $f_\theta(\cdot)$ is the combined transfer function of the transmitter components, and $\mathbf{x^{(n)}} = [x(n-L),\ldots, x(n), \ldots x(n+L)]^\top$ is the $2L+1$-long vector of input signals that contribute to $y(n)$. Note that $N$ is the number of samples over which the empirical MSE is calculated.
In practice, optimizing the parameters of  $f_\mathrm{DPD}(\cdot)$  is challenging because a differentiable form of $f_\theta(\cdot)$, which includes the transfer function of VCSEL, is typically unknown. Here, differentiable surrogate models of VCSELs based on Volterra or NNs developed in Sec. \ref{sec:modeling} find direct application. Alternatively, methods for gradient-free transmitter learning (see Sec.~\ref{sec:gradient-free}) can be used for the DPD optimization. The resulting approach, known as the direct learning architecture (DLA), uses various NN architectures such as FCNN\cite{hadi2021neural,minelli2022end}, LSTM\cite{paryanti2020direct} and CNN \cite{bajaj2020single}. However, DLA has the disadvantage that the performance of the learned DPD is highly dependent on the accuracy of the surrogate model. 

A different approach, namely the indirect learning architecture (ILA), estimates the DPD parameters in an indirect fashion by first learning the inverse response of the transmitter, which we refer to as the post-distorter $f^{-1}_\theta(\cdot)$. The post-distorter is then used as the pre-distorter. An example of the above mentioned DPD employing the ILA is depicted in Fig.~\ref{fig:ILA}, where the post-distorter is first learned by minimizing the MSE between the transmitter input $x(n)$ and the post-distorter output $\hat{x}(n)$, and then the post-distorter parameters are copied to the pre-distorter. The corresponding optimization problem can be summarized as
\begin{align}
    \hat{f}_\mathrm{DPD} \triangleq \hat{f}^{-1}_\theta =\underset{f^{-1}_\theta}{\arg \min}  \frac{1}{N}\sum_{n=1}^{N} |\hat{x}(n)- x(n)|^2,
\end{align}
where $\hat{x}(n) = f^{-1}_\theta(\mathbf{y}^{(n)})$ , $y(n) = f_\theta(\mathbf{x}^{(n)})$ and $\mathbf{y^{(n)}} = [y(n-L),\ldots, y(n), \ldots y(n+L)]^\top$. 

In general, it has been shown that DPD employing DLA can achieve better performance than that using ILA, in case a good surrogate transmitter model is available\cite{paryanti2020direct,bajaj2022deep}. However, it should be noted that DPDs employing ILA are more commonly used in practice due to their low implementation complexity. Therefore, it might serve as a suitable candidate for VSCEL-induced nonlinearity compensation.

\begin{table*}[]
    \centering
    \begin{tabular}{|c|c|p{0.15\linewidth}|p{0.4\linewidth}|}
    \hline
        \textbf{Paper} & \textbf{Application} & \textbf{NN architecture} & \textbf{Contribution} \\
        \hline
        \hline
        \cite{Karanov2018} & IM/DD systems & FCNN & Explored AE-based learning for fiber-optics\\
        \hline
        \cite{Karanov2019end} & IM/DD systems & BRNN &  Addressed memory effects\\
        \hline
        \cite{karanov2019deep} &IM/DD systems & BRNN &  Optimized bit-to-symbol mapping\\
        \hline
        \cite{Karanov2020} &IM/DD systems & FCNN &  Employed a GAN to obtain a model for an experimental IM/DD test-bed \\
        \hline
        \hline
        \cite{schaedler2020neural} & Coherent systems & FCNN &  Optimized both Amplified and Unamplified links\\
        \hline
        \cite{li2018achievable} & Coherent systems & FCNN &  Optimized for nonlinear phase noise channel models\\
        \hline
        \cite{Jones2018} & Coherent systems & FCNN &  Optimized for Gaussian noise (GN)-model and the nonlinear interference noise (NLIN)-model\\
        \hline
        \cite{jones2019end, Gumus2020} & Coherent systems & FCNN &  Optimized bit-to-symbol mapping\\
        \hline
        \cite{Uhlemann2020} & Coherent systems & FCNN & Used Split-step Fourier method (SSFM)\\
        \hline
        \cite{gaiarin2020end} & Coherent systems & NFT-NN &   Optimized nonlinear frequency division multiplexing system\\
        \hline
        \cite{Jovanovic2021} &Coherent systems & FCNN   & Cubature Kalman Filter based gradient-free approach\\
        \hline
        \cite{rode2022geometric} & Coherent systems & FCNN   & Optimized bit labeling including a differential blind phase search\\
        \hline
        \cite{jovanovic2022end} & Coherent systems & FCNN   & Robust to channel condition uncertainties\\
        \hline
        \hline
        \cite{Srinivasan2022learning} & VCSEL-based OIs & FCNN   & Optimized the PAM-4 levels\\
        \hline
        \cite{ Minelli2022nonlinear} & VCSEL-based OIs & FCNN   & Non-linear predistortion\\
        \hline
    \end{tabular}
    \caption{State of the art on AE-based E2E learning for fiber-optic communications, using different types of NN architectures, either at the transmitter or at the reviever, such as fully connected neural network (FCNN), bidirectional recurrent neural network (BRNN), generative adversarial network (GAN). NFT stands for nonlinear Fourier transform.} \vspace{-5mm}
    \label{tab:prior_ae}
\end{table*}

\section{End-to-End Learning Using Autoencoders} \label{sec:AutoEncoder}

The modular approach considered so far does not guarantee optimal E2E performance. To address this, the methods presented in the earlier section can be combined with E2E learning using AEs. 
In this section, after a literature survey, we review the general principles of AE-based communication systems and summarize the challenges of E2E learning in a VCSEL setting: (i) the need for \emph{differentiable channel models}; (ii) the lack of \emph{adaptivity to the temperature}. We show how a surrogate model (as defined earlier in Section \ref{sec:modeling}) can overcome the first challenge, while to address the second challenge, we elaborate on the temperature-adaptive transceiver to incorporate flexibility into our learning framework so that the transceiver can be utilized in different deployment scenarios. Finally, we review and describe gradient-free approaches in cases lacking a differentiable channel model. 
 
 \subsection{Survey of Autoencoders in Fiber-Optic Communications}
AE-based deep NN architecture was proposed initially in \cite{o2017introduction} in the context of wireless links. It was then adapted to fiber-optic systems initially in \cite{Karanov2018} for optimizing IM/DD links. Since feed-forward NN cannot address the memory effects, bidirectional RNN (BRNN) was adopted in \cite{Karanov2019end}. Further improvements were made through AE-based bit-to-symbol mapping in \cite{karanov2019deep, jones2019end, Gumus2020}. AE-based learning has been explored in coherent optical systems under various scenarios in papers such as \cite{Karanov2020, Uhlemann2020, Jones2018}. Furthermore, gradient-free AE-optimization was initially explored for fiber-optic communication in \cite{Jovanovic2021}. Recently, AE-based optimization of VCSEL-based OI, utilizing a surrogate NN, was explored in \cite{Srinivasan2022learning}. AE-based E2E learning is still in its early days of application to VCSEL-based OIs but has shown promising potential in optimizing PAM levels \cite{Minelli2022nonlinear}. A comprehensive state-of-the-art, along with the type of NN used, the type of applications (conventional IM/DD links, Coherent systems, and VCSEL-based OIs), and their primary contributions have been provided in Table \ref{tab:prior_ae}.  
 
\begin{figure*}[t!]
    \centering
          \includegraphics[scale=0.57]{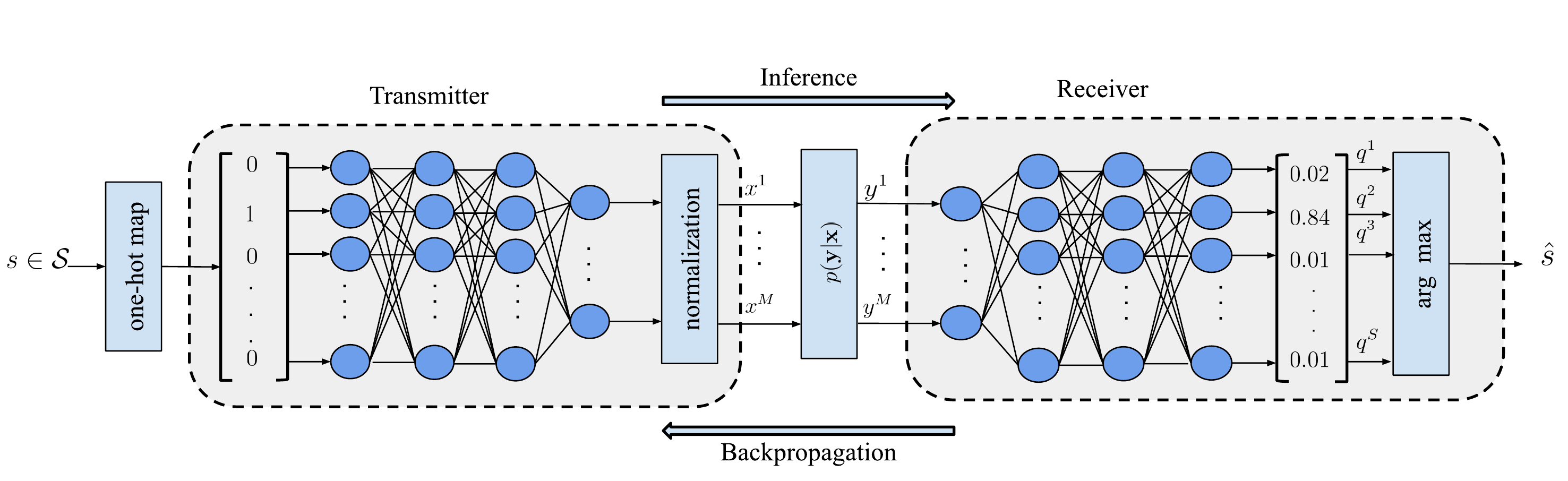}
          \caption{An AE-based communication system, where the transmitter and receiver are implemented by FCNNs.}
          \label{fig:general-ae}
\end{figure*}
\subsection{Conventional Autoencoder}\label{sec:conventional}
In an AE, the input to the encoder is reproduced at the output of the decoder---a principle on which communication systems are also modeled. Therefore, the encoder and decoder can be designed in such a way that they can replace the transmitter and receiver, respectively. The topology of a generic AE-based communication system is depicted in Fig. \ref{fig:general-ae}. In the AE structure, a message $s$ $\in$ $\mathcal{S}$ is first encoded into an $S$-dimensional one-hot vector, where the $s$-th element is $1$ and all the other elements are $0$. Here, $\mathcal{S}$ is the message set and $\mathcal{S} = \{1, \hdots, S\}$. The transmitter NN can be viewed as a function that computes a $M$-dimensional representation $\bm x$ for every one-hot representation  $s$. In other words, $\bm x = f_\theta(s)$, where $f_\theta$ is the function denoting the transmitter NN. Here, $M$  denotes the number of real/complex channel uses. The transmit power constraint is enforced by a normalization layer.

The symbol $\bm x$ is sent over the channel, whose channel law can be denoted by $p\left(\bm y| \bm x\right)$ in $M$ channel uses, and the symbol observed at the receiver is $\bm y$. The receiver NN can be viewed as a function that computes an $S$-dimensional posterior probability $\bm q$. The receiver then estimates the transmitted message $\hat s= {\arg\max_s}\left[\bm q\right]_s$, where $\left[\bm x\right]_s$ returns the $s$-th element of $\bm x$.

To optimize the transmitter and receiver parameters, $\theta$ and $\phi$, we use the cross-entropy loss
\begin{equation}
    \mathcal{L}\left(\theta, \phi\right) = -\mathbb{E}\left\{\int  \log\left[ f_{\phi}\left( \bm y\right)\right]_{s} p(\bm y|f_\theta(s))
    \mathrm{d}\bm y 
    \right\},
\end{equation}
where $\mathbb{E}\left\{.\right\}$ is the expectation operator over $s$. The optimization is performed iteratively, where in each training iteration $t$,
the transmitter maps a minibatch of $ B_t$ randomly chosen training examples $s_k$ to symbols $\bm x_k=f_\theta(s_k)$ for $k = \left\{1, \hdots,B_t\right\}$ and then sends them  through the channel. The receiver maps the observations $\bm y_k$ to the probability vectors $f_\phi\left(\bm y_k\right)$ for $k = \left\{1, \hdots, B_t\right\}$ . Finally, the empirical cross-entropy loss associated with the training examples is calculated.

After the training procedure, the BER (or any other relevant performance metric) is measured during testing. 
Backpropagation relies on calculating the gradient of the loss function with respect to all the trainable parameters of the receiver and the transmitter. Therefore, if any component has a non-differentiable/intractable model, the encoder will be isolated from the rest of the computational chain, and the gradient of the loss function with respect to the encoder weights cannot be calculated. 
At this juncture, the importance of the surrogate models in the E2E structure has to be reiterated. Hence, differentiable surrogate models, especially the NN models, that capture the dynamics of VCSELs and accurately reproduce the optical waveforms generated by the VCSELs under different operating temperatures are discussed in detail in Section \ref{sec:modeling}. Alternatively, some methods for gradient-free transmitter optimization will be reviewed below in Sec.~\ref{sec:gradient-free}. The NN surrogate models can then be incorporated into an AE that models an E2E fiber-optic system and trained to improve performance. 
\begin{figure}[t!]
    \centering
            \includegraphics[scale=0.48]{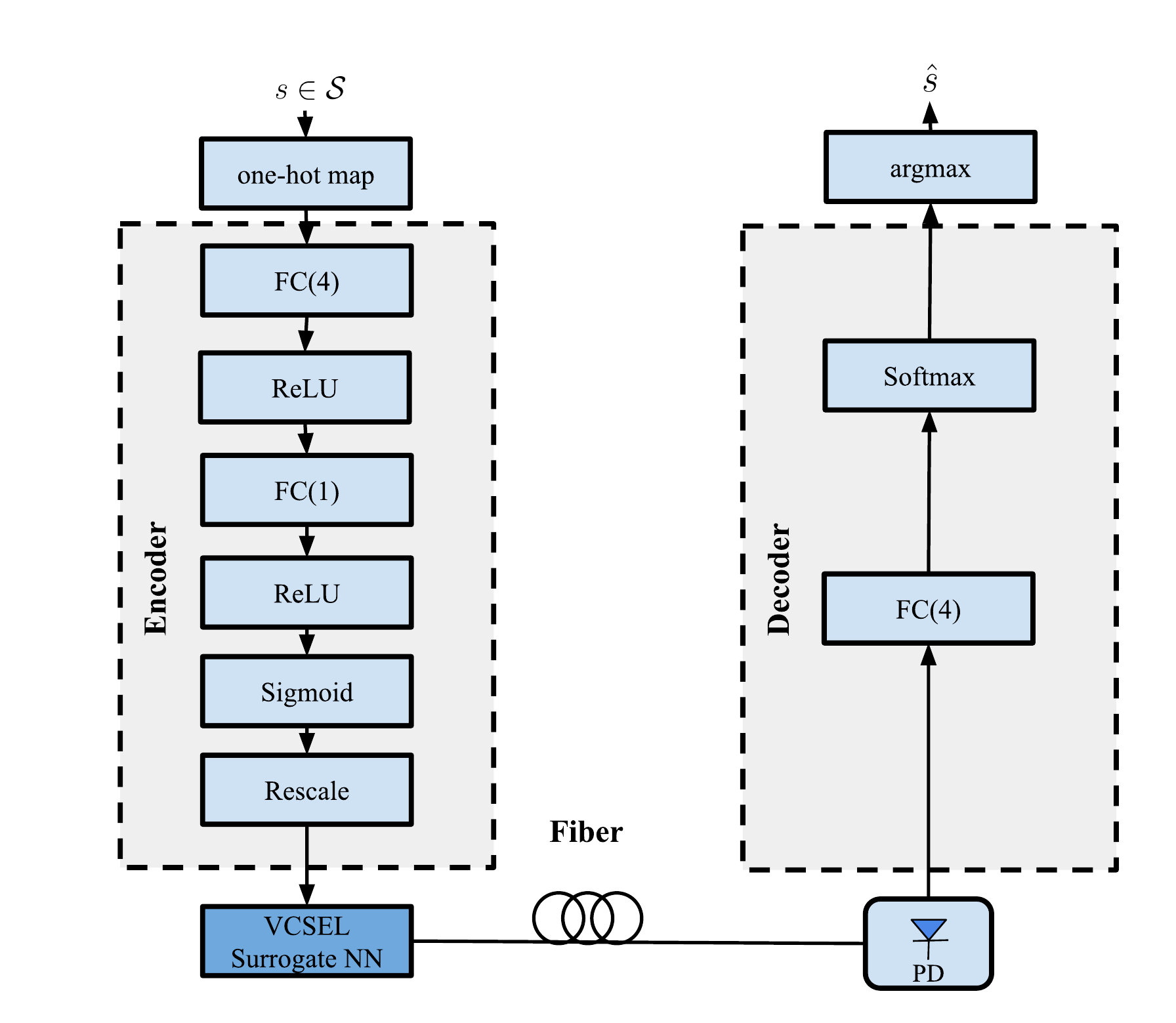}
            \caption{AE structure containing two encoder layers, the surrogate NN for VCSEL and a single decoder layer. FC denotes fully connected layers and ReLU denotes rectified linear units (ReLUs). The numbers in the parentheses represent the number of neurons in a specific layer. PD denotes photodiode. Like VCSEL, the fiber and PD are replaced by surrogate models for backpropogation.}
            \label{fig:ecoc-ae}
\vspace{-10pt} \end{figure}

\begin{figure}[t!]
\centering
%
%
\begin{tikzpicture}

\begin{axis}[%
width=2.528in,
height=1.954in,
at={(0.5in,0.5in)},
scale only axis,
xmin=6,
xmax=16,
xlabel style={font=\color{white!15!black}},
xlabel={SNR (dB)},
ymode=log,
ymin=1e-06,
ymax=1,
yminorticks=true,
ylabel style={font=\color{white!15!black}},
ylabel={BER},
axis background/.style={fill=white},
xmajorgrids,
ymajorgrids,
legend cell align={left},
legend style={at={(0.0,0.0)}, nodes={scale=0.94, transform shape},anchor=south west, align=left, draw=white!15!black}
]
\addplot [color=blue, line width=1.5pt, mark=asterisk, mark options={solid, blue}]
  table[row sep=crcr]{%
6	0.195159756437808\\
7	0.132145689311204\\
8	0.0783312049808448\\
9	0.0404478640307763\\
10	0.0180901628114653\\
11	0.00714806433257899\\
12	0.00235852122669104\\
13	0.0006000054000486\\
14	0.000116501048509437\\
15	1.65001485013365e-05\\
16	2.000009000081e-06\\
17	0\\
18	0\\
};
\addlegendentry{$\text{Equidistant 5}^\circ\text{C}$}

\addplot [color=blue, dashed, line width=1.5pt, mark=o, mark options={solid, blue}]
  table[row sep=crcr]{%
6	0.192433231899087\\
7	0.127879150912358\\
8	0.0733106597959382\\
9	0.0351213160918448\\
10	0.0140641265771392\\
11	0.00454004086036774\\
12	0.00114401029609266\\
13	0.000221501993517942\\
14	2.8000252002268e-05\\
15	3.5000315002835e-06\\
16	3e-07\\
17	5.000045000405e-07\\
18	5.000045000405e-07\\
};
\addlegendentry{$\text{AutoEncoder 5}^\circ\text{C}$}

\addplot [color=red, line width=1.5pt, mark=asterisk, mark options={solid, red}]
  table[row sep=crcr]{%
6	0.196412535425638\\
7	0.135109431969775\\
8	0.0823884829926939\\
9	0.0447938062885132\\
10	0.0219953959171265\\
11	0.0100471808492553\\
12	0.00413107435933847\\
13	0.0014580262444724\\
14	0.000466008388150987\\
15	0.000110001980035641\\
16	2.40004320077761e-05\\
17	1.00001800032401e-06\\
18	1.00001800032401e-06\\
};
\addlegendentry{$\text{Equidistant 95}^\circ\text{C}$}

\addplot [color=red, dashed, line width=1.5pt, mark=o, mark options={solid, red}]
  table[row sep=crcr]{%
6	0.182504285077131\\
7	0.120836175051151\\
8	0.0693772487904782\\
9	0.033623605224894\\
10	0.0139252506545118\\
11	0.00466008388150987\\
12	0.00139202505645102\\
13	0.00031700570610271\\
14	3.70006660119882e-05\\
15	5.00009000162003e-06\\
16	4e-07\\
17	0\\
18	0\\
};
\addlegendentry{$\text{AutoEncoder 95}^\circ\text{C}$},
\end{axis}
\end{tikzpicture}%
\caption{BER vs SNR comparison for AE and an NN-decoder that decodes equidistant PAM levels.}
\label{fig:bervssnrforae}
\vspace{-10pt} \end{figure}
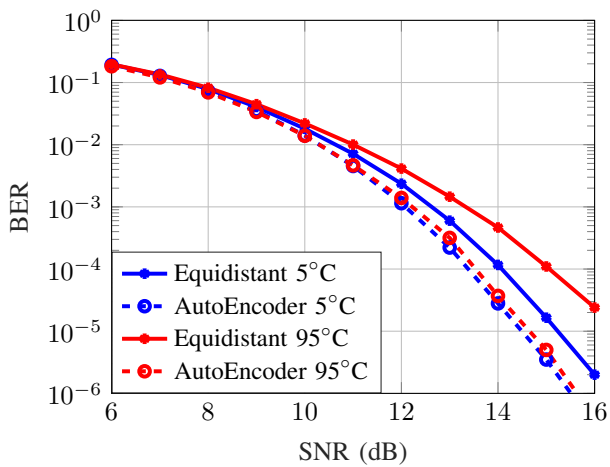

\begin{example}[AE with surrogate model]
One such surrogate model for AE is shown in Fig.~\ref{fig:ecoc-ae}, in which the VCSEL is replaced with a surrogate NN. The remaining components (PIN and fiber) are easily modeled with differentiable functions. The AE learns to explore new geometrically shaped constellations for different temperatures. The BER vs. signal-to-noise ratio (SNR) results of the AE and the equidistant-decoder (ED) are plotted in Fig. \ref{fig:bervssnrforae}. At $95\degree$ C, the sensitivity improvement is about $1.5$ dB, validating the idea that equidistant PAM levels do not attain an optimal BER. At $5\degree$ C, the sensitivity improvement is about $0.5$ dB.
\end{example}

\subsection{Temperature-adaptive AE}

From Fig.~\ref{fig:bervssnrforae}, it is clear that 
optimal modulation formats, transmit waveforms, and receiver processing algorithms may depend heavily on temperature and should be adapted accordingly. 
This section will review several approaches for incorporating adaptivity into the ML models of the E2E learning framework. 



\subsubsection{Explicit Temperature-dependent Learning}
A straightforward approach for dealing with adaptivity is to train a dedicated model for each deployment scenario, e.g., one model per temperature. However, this becomes infeasible if the number of scenarios is large or infinite. A commonly used trick to circumvent the computational burden associated with training and storing a large number of NNs is to train a family of network parameters that are themselves parameterized. A simple way to accomplish this is to append the external system parameter, e.g., the temperature $T$, as an additional input to the NN via concatenation with the regular input $\boldsymbol{s}$. 
A more dynamic approach is via the use of a second auxiliary NN. More precisely, if $f_\rho(\boldsymbol{s})$ denotes the original NN mapping, one may generate the parameter vector $\rho$ (i.e., the weights and biases) using another NN according to $\rho = f_\omega(T)$, where $\omega$ are the new trainable parameters. Such approaches have previously been studied in the context of providing SNR-adaptive forward error correction schemes; see \cite{Lian2019isit} for more details. 

\subsubsection{Robust Learning}
Another approach for providing ML models with some form of robustness to temperature variation is to simultaneously train them on data from a variety of different deployment scenarios. For example, the authors in \cite{Karanov2018} propose to train their AE system on data corresponding to a range of different fiber lengths, which then offers some flexibility if the exact transmission distance is unknown in advance. Applying a similar approach to VCSEL-based E2E learning could lead to \emph{temperature-agnostic} models that, once trained, can be expected to work well over a range of different temperatures. However, this approach effectively leads to ``compromise'' solutions, which may be heavily suboptimal given any fixed temperature.

\subsubsection{Online Learning}
While the above approaches are relatively easy to implement, they ultimately hinge on the assumption that the data used for ``offline" training accurately reflect the "online" deployment conditions. If this is not the case, one may argue that a better approach to adaptive E2E learning is via in-situ training of all NNs using live, real-time training data from the actual transmissions. On the receiver side, this generalizes the adaptive linear equalizers commonplace in optical receivers for dealing with time-varying impairments. However, a real-time adaptation of nonlinear NN equalizers would require some form of online gradient backpropagation, which may be challenging to implement in hardware for VCSEL-based interconnects, given the severe complexity constraints. Moreover, an in-situ adaptation of the transmitter NN is even more challenging due to the absence of a differentiable channel model required for the gradient computation. 
Also, to limit the computational burden associated with NN training, some recent work focuses on meta-learning \cite{Park2020, Simeone2020}. Here, the main idea is to learn model initialization that can facilitate faster retraining, which can also be exploited in combination with decision-directed training \cite{Shlezinger2020}.

\begin{figure}[t!]
\centering
\includegraphics[scale=0.3]{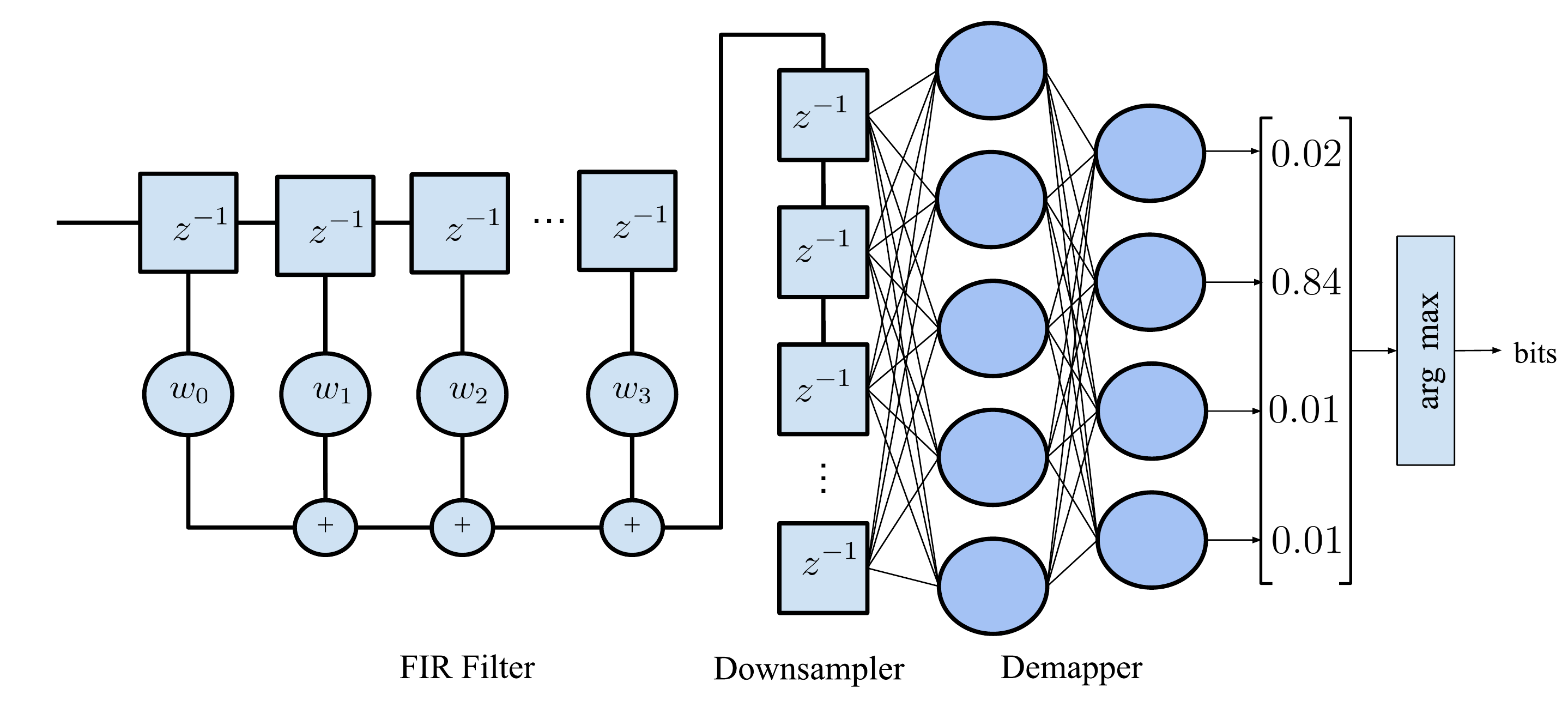}
\caption{Trainable receiver architecture that consists of FIR filter and TDNN that implements downsampling and de-mapping.}
\label{fig:trainablerx}
\vspace{-10pt} \end{figure}

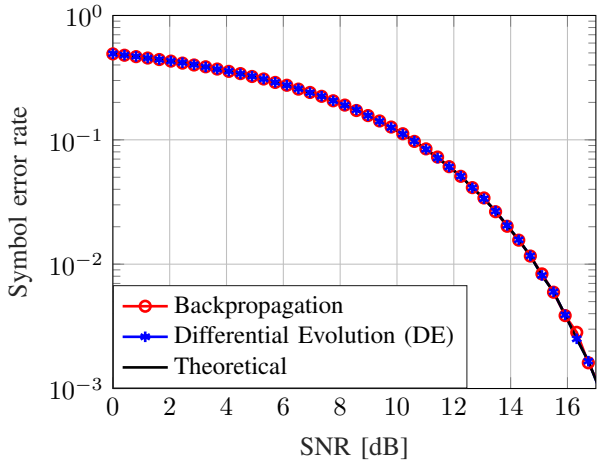
\begin{figure}[t!]
\centering
%
%
\begin{tikzpicture}

\begin{axis}[%
width=2.528in,
height=1.954in,
at={(0.5in,0.5in)},
scale only axis,
xmin=0,
xmax=17,
xlabel style={font=\color{white!15!black}},
xlabel={SNR [dB]},
ymode=log,
ymin=0.001,
ymax=1,
yminorticks=true,
ylabel style={font=\color{white!15!black}},
ylabel={Symbol error rate},
axis background/.style={fill=white},
xmajorgrids,
ymajorgrids,
legend cell align={left},
legend style={at={(0.0,0.0)}, nodes={scale=0.94, transform shape},anchor=south west, align=left, draw=white!15!black}
]
\addplot [color=red, line width=1.0pt, mark=o, mark options={solid, red}]
  table[row sep=crcr]{%
0	0.4911\\
0.40816	0.47893\\
0.81633	0.46666\\
1.22449	0.45478\\
1.63265	0.44236\\
2.04082	0.42953\\
2.44898	0.41392\\
2.85714	0.39987\\
3.26531	0.38671\\
3.67347	0.37086\\
4.08163	0.35522\\
4.4898	0.34088\\
4.89796	0.32316\\
5.30612	0.3085\\
5.71429	0.28945\\
6.12245	0.27467\\
6.53061	0.25571\\
6.93878	0.24041\\
7.34694	0.22379\\
7.7551	0.20602\\
8.16327	0.19025\\
8.57143	0.1721\\
8.97959	0.15668\\
9.38776	0.1418\\
9.79592	0.12668\\
10.20408	0.11189\\
10.61224	0.09711\\
11.02041	0.08458\\
11.42857	0.0727\\
11.83673	0.06084\\
12.2449	0.05087\\
12.65306	0.04127\\
13.06122	0.03407\\
13.46939	0.02642\\
13.87755	0.02018\\
14.28571	0.0156\\
14.69388	0.01162\\
15.10204	0.00835\\
15.5102	0.00595\\
15.91837	0.00386\\
16.32653	0.00283\\
16.73469	0.00161\\
17.14286	0.00104\\
17.55102	0.00061\\
17.95918	0.00025\\
18.36735	0.00019\\
18.77551	8e-05\\
19.18367	3e-05\\
19.59184	1e-05\\
20	1e-05\\
};
\addlegendentry{Backpropagation}

\addplot [color=blue, line width=1.0pt, mark=asterisk, mark options={solid, blue}]
  table[row sep=crcr]{%
0	0.49214\\
0.40816	0.47987\\
0.81633	0.46839\\
1.22449	0.45434\\
1.63265	0.4422\\
2.04082	0.42987\\
2.44898	0.41714\\
2.85714	0.40173\\
3.26531	0.38564\\
3.67347	0.37001\\
4.08163	0.35663\\
4.4898	0.3397\\
4.89796	0.32475\\
5.30612	0.30835\\
5.71429	0.29013\\
6.12245	0.27493\\
6.53061	0.25696\\
6.93878	0.23949\\
7.34694	0.22434\\
7.7551	0.20599\\
8.16327	0.18999\\
8.57143	0.17414\\
8.97959	0.1565\\
9.38776	0.14094\\
9.79592	0.12558\\
10.20408	0.1113\\
10.61224	0.09767\\
11.02041	0.084\\
11.42857	0.07147\\
11.83673	0.06054\\
12.2449	0.05104\\
12.65306	0.04112\\
13.06122	0.03362\\
13.46939	0.02672\\
13.87755	0.02048\\
14.28571	0.0157\\
14.69388	0.01166\\
15.10204	0.00812\\
15.5102	0.00599\\
15.91837	0.00391\\
16.32653	0.00251\\
16.73469	0.00166\\
17.14286	0.00097\\
17.55102	0.00053\\
17.95918	0.00032\\
18.36735	0.00013\\
18.77551	0.0001\\
19.18367	3e-05\\
19.59184	3e-05\\
20	0\\
};
\addlegendentry{Differential Evolution (DE)}

\addplot [color=black, line width=1.0pt]
  table[row sep=crcr]{%
0	0.49214\\
0.40816	0.47987\\
0.81633	0.46839\\
1.22449	0.45434\\
1.63265	0.4422\\
2.04082	0.42987\\
2.44898	0.41714\\
2.85714	0.40173\\
3.26531	0.38564\\
3.67347	0.37001\\
4.08163	0.35663\\
4.4898	0.3397\\
4.89796	0.32475\\
5.30612	0.30835\\
5.71429	0.29013\\
6.12245	0.27493\\
6.53061	0.25696\\
6.93878	0.23949\\
7.34694	0.22434\\
7.7551	0.20599\\
8.16327	0.18999\\
8.57143	0.17414\\
8.97959	0.1565\\
9.38776	0.14094\\
9.79592	0.12558\\
10.20408	0.1113\\
10.61224	0.09767\\
11.02041	0.084\\
11.42857	0.07147\\
11.83673	0.06054\\
12.2449	0.05104\\
12.65306	0.04112\\
13.06122	0.03362\\
13.46939	0.02672\\
13.87755	0.02048\\
14.28571	0.0157\\
14.69388	0.01166\\
15.10204	0.00812\\
15.5102	0.00599\\
15.91837	0.00391\\
16.32653	0.00251\\
16.73469	0.00166\\
17.14286	0.00097\\
17.55102	0.00053\\
17.95918	0.00032\\
18.36735	0.00013\\
18.77551	0.0001\\
19.18367	3e-05\\
19.59184	3e-05\\
20	0\\
};
\addlegendentry{Theoretical}
\end{axis}

\end{tikzpicture}%
\caption{Symbol error rate as a function of signal-to-noise-ratio for the receiver architecture trained by differential evolution and backpropagation.}
\label{fig:SERvsBP_DE}
\vspace{-10pt} \end{figure}

\subsection{Gradient-free Approaches}
\label{sec:gradient-free}
All of the E2E learning techniques discussed in the previous subsections used differentiable models to optimize the AE using backpropagation. An accurate differentiable model that captures the effects of VCSEL is indeed challenging to develop, as seen from Section \ref{sec:modeling}, and leads to a model discrepancy. This could be true for any other component in the chain shown in Fig. \ref{fig:E2E-opticalinterconnects}. Furthermore, training of the transmitter and receiver NN architectures may not be feasible using backpropagation for the experimental or hardware implementation of the system. A solution to the above-mentioned problem is to perform training using gradient-free optimization methods. Numerous algorithms perform gradient-free optimization; however, most rely on various approximations of the gradient \cite{raj2018backpropagating}. In that respect, at their best, the performance will approach the performance of the propagation. It is, therefore, always advisable to use backpropagation for training when possible. 

Solutions based on the Cubature Kalman Filter (CKF) have been recently demonstrated for training NN transmitter, and receiver architectures \cite{Jovanovic2021}. The advantage of training NNs using this method is that the entire channel model can be easily embedded into the optimization. The disadvantage is that it requires matrix inversion, which is time-consuming. Moreover, especially at the beginning of the training process, the matrix may be singular and thus non-invertible. 

Another approach for gradient-free training of transmitter and receiver architectures is to employ evolutionary algorithms. An advantage of evolutionary algorithms is that the rate of exploration-exploitation can be controlled, which may be helpful to avoid getting trapped in local minima of the loss function. One of the challenges with evolutionary algorithms is that there are many parameters to tune. However, an evolutionary algorithm employing differential evolution has only two hyperparameters, which are easily tuneable. It has been shown that differential evolution (DE) is highly effective in performing online experimental optimization of flatness of optical frequency combs \cite{pinto2021optimization}. 

\begin{example}[AE with differential evolution]
A sample trainable receiver NN architecture that consists of an FIR filter, a down-sampler, and a demapper, implemented as a TDNN, and uses DE for optimization, is shown in Fig.~\ref{fig:trainablerx} as a proof-of-concept. The objective of the optimization is to find a set of FIR filter coefficients and neural-networks weights that minimize the loss function, implemented as the cross-entropy. The channel under consideration is AWGN. 
In Fig.~\ref{fig:SERvsBP_DE}, the symbol error rate is plotted as a function of SNR for the receiver architecture shown in Fig.~\ref{fig:trainablerx} trained by the differential evolution and backpropagation. It is observed that the receiver performance obtained by training the receiver using differential evolution and backpropagation overlaps with the theoretical curve. A similar gradient-free approach can also be applied to a VCSEL-based OI system. 
\end{example}

\section{Conclusions and Outlook} \label{sec:Conclusions}
Besides the approaches mentioned above that can result in impressive benefits in designing an OI, there are several opportunities to explore in the context of E2E learning. For example, as an improvement over feed-forward NN, bidirectional recurrent neural networks (BRNN), which have been used in conventional IM/DD links, can be adopted for VCSEL-based OIs. Since BRNN can process sequential data using internal states, they can compensate memory effects of VCSELs, and the fiber \cite{Karanov2019end}. Also, one can explore an AE that can optimize the bit mapping jointly with the position of the constellation points using a loss function based on the bit-wise or generalized mutual information (GMI) \cite{jones2019end, Gumus2020}. Since the optimization landscape in such a case is highly non-convex, a good strategy is to repeat the optimization with multiple starting points to find the global optimum. Alternatively, initialization to known Gray-labed constellations, cyclical learning rates, or the binary switching algorithm have been shown to achieve more reliable results compared to random initializations \cite{Gumus2020}. Furthermore, the issue of modeling the experimental environment can be overcome by generating the channel model including the VCSEL's by a generative adversarial network (GAN). Gradient backpropagation can then be used to train the generative model on experimental data and use it to optimize the transceiver \cite{Karanov2020}. Finally, it is possible to explore ML-based forward error correction (FEC) codes that can compensate for
decision errors in the E2E learning framework \cite{balevi2020autoencoder}.


\end{document}